\documentclass[useAMS,usenatbib]{mn2e}
\usepackage[pdftex]{graphicx}
\pdfoutput=1
\usepackage{amssymb,amsmath,natbib}
\title[RADAMESH]{RADAMESH: Cosmological Radiative Transfer for Adaptive Mesh Refinement Simulations}
\author[S. Cantalupo \& C. Porciani]{Sebastiano Cantalupo$^1$\thanks{E-mail:cantal@ast.cam.ac.uk} and Cristiano Porciani $^2$\\
$^1$ Kavli Institute for Cosmology, Cambridge \& Institute of Astronomy, Madingley Road, Cambridge CB3 0HA, UK\\
$^2$ Argelander Institute f\"ur Astronomie, Auf dem H\"ugel 71, Bonn, Germany}

\begin{document}

\date{Accepted 2010 September 30. Received 2010 September 30; in original form 2010 June 28}
\pagerange{\pageref{firstpage}--\pageref{lastpage}} \pubyear{}

\maketitle

\label{firstpage}

\newcommand{\cnamens}{\texttt{RADAMESH}}
\newcommand{\cname}{\cnamens\ }

\begin{abstract}

We present a new three-dimensional radiative transfer (RT) code, \cnamens,
based on a ray-tracing, photon-conserving and adaptive (in space and time)
scheme. \cname uses a novel Monte Carlo approach to sample the radiation field
within the computational domain on a ``cell-by-cell'' basis. Thanks to this algorithm,
the computational efforts are now focused where actually needed, i.e.
within the Ionization-fronts (I-fronts). This results in an increased accuracy level and,
at the same time, a huge gain in computational speed with respect to a
``classical'' Monte Carlo RT, especially when combined with
an Adaptive Mesh Refinement (AMR) scheme. Among several new features, \cname is able to adaptively refine the 
computational mesh in correspondence of the I-fronts, allowing to fully resolve them
within large, cosmological boxes. We follow the propagation of ionizing
radiation from an arbitrary number of sources and from the recombination radiation
produced by H and He. The chemical state 
of six species (HI, HII, HeI, HeII, HeIII, e) and gas temperatures are
computed with a time-dependent, non-equilibrium chemistry solver.   
We present several validating tests of the code, including the standard
tests from the RT Code Comparison Project and a new set of tests aimed at substantiating
the new characteristics of \cnamens. Using our AMR scheme, we show that properly resolving the I-front
of a bright quasar during Reionization produces a large increase of the predicted gas temperature within the whole HII region. 
Also, we discuss how H and He recombination radiation
is able to substantially change the ionization state of both species
(for the classical Str\"omgren sphere test) with respect to the widely used 
``on-the-spot'' approximation. 

\end{abstract}


\newcommand{\mrm}{\mathrm}
\newcommand{\HI}{\mathrm{HI}}
\newcommand{\HII}{\mathrm{HII}}
\newcommand{\HeI}{\mathrm{HeI}}
\newcommand{\HeII}{\mathrm{HeII}}
\newcommand{\HeIII}{\mathrm{HeIII}}

\section{Introduction}

The interaction between matter and radiation is one of the fundamental mechanisms shaping the
distribution of the baryonic component in the Universe, from stellar to cosmological scales. 
This process often couples wildly different scales and the large dynamical 
range makes accurate modelling difficult.
A notable example is given by the reionization of cosmic hydrogen 
at redshift $6\lesssim z \lesssim 12$ (for a review, see, e.g., Meiksin 2009).  
During this epoch,
the HII regions generated by the first ionizing sources expand to intergalactic scales 
and overlap, leaving most of the Universe highly ionized and re-heated by several thousand Kelvin degrees. 
The brightest, high-redshift quasars, 
may have produced even larger HII regions before the end of Reionization, 
with linear sizes extending up to a hundred comoving Mpc. Nonetheless, most of the
evolution in the gas temperature and ionization state still happened on much
smaller scales, i.e. within the Ionization-fronts, whose sizes are
comparable to the local photon mean-free-path just outside the HII regions, 
about three orders of magnitude smaller than the HII regions themselves for gas
at mean cosmic density. 

Even without considering such an extreme case in dynamical range, the numerical solution of the
full radiative transfer equations still represents a big computational challenge. This is mostly
due to two factors. The first is the high-dimensionality: seven variables are required
for a full specification of the radiation field (three spatial variables,
two angular directions, photon frequency, and time). The second is the intrinsic ``non-locality'' of the
problem: 
the radiation field at a given point is determined by the gas properties along the lines of sight towards
all the sources of radiation (including the diffuse medium itself).

For these reasons, current numerical models rely on approximations aimed to decrease the problem
dimensionality and, at the same time, the computational costs associated with the ``non-locality'' of the
full radiation transfer (e.g., Abel, Norman \& Madau 1999, Gnedin \& Abel 2001, Maselli, Ferrara \& Ciardi 2003,
Razoumov \& Cardall 2005, Mellema et al. 2006, Rijkhorst et al. 2006, Ritzerveld \& Icke 2006, Whalen \& Norman 2006;
for a review, see also Iliev et al. 2006 and references therein; more recent codes include, e.g., Trac \& Cen 2007, 
Semelin et al. 2007, Aubert \& Teyssier 2008, Altay et al. 2008, Pawlik \& Schaye 2008, Finlator et al. 2009, Petkova \& Springel 2009).
A widely used approximation is the so called ``method of characteristics'' which is particularly suited for cases where the 
radiation field is completely dominated by individual sources (with no contribution from the radiative 
recombinations produced by the medium itself). In this scheme, the radiation field is determined
simply from the column densities along a line of sight between the sources and a given point in the computational
domain, discretized in cells (or particles). 
In the ``long-characteristics'' flavour, the column densities are directly calculated with a single ray
from the sources to the cell centre, summing up all the contributions cut through the cells in between.
In the faster, but less accurate, ``short-characteristics'' version, the column densities are determined in an ordered
way starting from the cells closer to the source and moving outwards after summing up (using some interpolation scheme) 
previous contributions. 
In both cases, sampling the radiation field with a single ray from the sources to the cell centre 
may be problematic in particular situations, 
i.e. when the cells are optically thick and in presence of
strong density gradients. Practically, the consequences may be the loss of photon conservation and numerical artifacts in the
I-front shapes (see, e.g., Mellema et al. 2006).

 In a cosmological situation, e.g. for the reionization scenario described before, we would like to have an accurate method
for the very typical case in which numerical resolution imposes optically thick cells.
This may be obtained sampling the radiation field in a statistical, homogeneous way, using the widely used Monte Carlo
techniques. In this case, a series of rays is randomly casted around the sources in order to obtain the right
solid angle distribution, averaging the contributions to the radiation field within individual (three-dimensional) volume elements.
The drawbacks of this method are the computational costs: many rays (per cell) are required to avoid statistical noise and 
the method becomes less and less efficient as one moves outwards from the sources (but see, Abel \& Wandelt 2002). 
Moreover, if the grid is composed
by cells of different sizes, the required number of rays to be casted is determined by the size of the 
smallest cells, irrespectively of the volume fraction they actually occupy.

 In this work, we develop a novel approach aimed to obtain an accurate solution of the radiative transfer problem
in cosmological situations in an efficient way: combining
the Monte Carlo scheme with the ``cell-by-cell'' approach typical of the characteristic methods.
As we mentioned at the beginning of this section, despite the large scales associated with, e.g., the reionization
process, the medium properties are very often determined within a small fraction of its volume, i.e.
within the I-fronts. The basic idea is to concentrate the computational efforts onto this part of the
computational volume, where they are actually needed. This is achieved thanks to a new algorithm which casts, 
with the correct solid angle distribution, a series of rays for each cell \emph{individually}. 
In this way, we are able to use an algorithm that adaptively determines: 
i) the number of rays needed for a particular cell to achieve the convergence of the
radiation field (irrespectively of the cell size, or distance from the source), ii) in which part of the
volume to apply the (expensive) Monte Carlo ray-tracing. The result is that the computing time of the RT scales now with the
number of cells to be evolved during a simulation-step, i.e., typically the cells contained within the I-fronts, with
a huge gain in computational speed (and accuracy) with respect to a classical Monte Carlo. 

 With the new ``cell-by-cell'' approach we are also able to gain the full advantages of Adaptive Mesh Refinement (AMR)
methods (see, e.g. Berger \& Oliger 1984) that increase the spatial resolution where needed. 
In particular, we implemented a scheme, based on the original clustering algorithm by Berger \& Rigoutsos (1991), to adaptively refine the mesh in correspondence
of the I-fronts during the course of the simulation. This allows us to fully resolve the small scales associated with the
I-fronts within the large, cosmological simulation boxes that follow the reionization process. Resolving the I-fronts
has a profound impacts for a large range of astrophysical applications, e.g., to accurately predict the temperature
state of the gas surrounding a bright quasar during Reionization. In the validating test section (section 4), we
will present an example of how resolving the I-front may be important in the determination of the gas temperature
within the \emph{whole} HII region. 

The paper is organized as follows. In section 2, we review the basic radiative transfer equation. 
The computational method used by our new RT code, \cname (\textbf{R}adiative-transfer on \textbf{ADA}ptive \textbf{MESH}),
is presented in section 3, In section 4, we show the
validating tests of the code. We conclude in section 5.

\section{Basic equations}

The cosmological radiative transfer equation in comoving coordinates (e.g., Norman, Paschos \& Abel 1998) is given by

\begin{equation}
\frac{1}{c} \frac{\partial I_{\nu}}{\partial t} + \frac{\hat{n} \cdot
\nabla I_{\nu}}{\bar{a}} - \frac{H(t)}{c} (\nu \frac{\partial I_{\nu}}
{\partial \nu} - 3 I_{\nu}) = k_{\nu}(S_{\nu}-I_{\nu})\ , 
\label{generalrteq}
\end{equation}
where $I_{\nu} \equiv I(t, \vec{x}, \hat{n}, \nu)$ is the
monochromatic specific intensity of the radiation field, $\hat{n}$ is
a unit vector along the direction of propagation of the ray, $H(t)
\equiv \dot{a}/a$ is the (time-dependent) Hubble constant, 
$\bar{a} \equiv \frac{1+z_{\mathrm{em}}}{1+z}$ is the ratio of cosmic scale
factors between photon emission at frequency $\nu$ and the present
time t, $k_{\nu}$ denotes the opacity at frequency $\nu$ and $S_{\nu}$
is the source function of the medium.

 If the scale of interest $L$ is much smaller than the Hubble radius, $c/H$, (as 
always in our case) and the medium properties are changing on a
time scale shorter than the light crossing time $L/c$, equation (\ref{generalrteq})
reduces to the classical, static radiative transfer equation:

\begin{equation}
\hat{n} \cdot\nabla I_{\nu} =  k_{\nu}(S_{\nu}-I_{\nu})\ .
\label{rteq}
\end{equation}
 This equation admits the following solution, 
\begin{equation}
\begin{split}
I_{\nu}(\tau_{\nu},\hat{n}) = I_{\nu}(0,\hat{n})\cdot e^{-\tau_{\nu}}+
\int_0^{\tau_{\nu}}S_{\nu}(\tau_{\nu},\hat{n})d\tau_{\nu}'\ ,
\end{split}
\label{Ieq}
\end{equation}
where $\tau_{\nu}$ is the optical depth along $\hat{n}$. This solution
is in general adequate for cosmological simulations except on small
distances from the sources (or, equivalently, for very bright sources).
In this case, the approximations break down allowing the Ionization-fronts
to expand faster than the speed of light.  
In section \ref{gammasec} we discuss an
approximate correction to solve this unphysical behaviour.
 It is often convenient to express $I_{\nu}$ as a sum of the attenuated, direct
 radiation from individual sources ($I_{\nu}^{\mrm{dir}}$) and the diffuse radiation ($I_{\nu}^{\mrm{diff}}$) 
 generated within the medium:
\begin{equation}
 I_{\nu}(\tau_{\nu},\hat{n}) = I_{\nu}^{\mrm{dir}}(\tau_{\nu},\hat{n})+I_{\nu}^{\mrm{diff}}(\tau_{\nu},\hat{n})\;. \\
 \label{Isplit}
\end{equation}

Using equation (\ref{Ieq}), the radiation field intensity can be specified
at any point of the simulation box given the value of the optical depth to the individual sources
and the source term of the medium. From the radiation field, it is then possible to derive for each species
$i$ the photoionization rate (per particle):
\begin{equation}
\Gamma_{i}(\vec{x})=\int_{4\pi} \int_{\nu_{i}}^{\infty} \sigma_{i}(\nu)\frac{I_{\nu}(\vec{x},\hat{n})}{h\nu}\mathrm{d}\nu\mathrm{d}\Omega\ 
\end{equation}
and the gas photoheating rate (per unit volume):
\begin{equation}
G_{i}(\vec{x})=n_{i}\int_{4\pi} \int_{\nu_{i}}^{\infty} \sigma_{i}(\nu)\frac{I_{\nu}(\vec{x},\hat{n})}{h\nu} h(\nu-\nu_{i})\mathrm{d}\nu\mathrm{d}\Omega\ ,
\end{equation}

where $\nu_{i}$ and $\sigma_{i}$ are the frequency threshold and the cross-section 
for the ionization of species $i$, respectively, and $n_{i}$ is the physical number
density. Analogously to eq. (\ref{Isplit}), we can split $\Gamma_i$ and $G_i$ 
into the sum of direct ($\Gamma_i^{\mrm{dir}}$,$G_i^{\mrm{dir}}$) and diffuse ($\Gamma_i^{\mrm{diff}}$,$G_i^{\mrm{diff}}$)
components.

Finally, $\Gamma_{i}$ and $G_{i}$ are used to compute the 
chemistry evolution of the neutral fraction of hydrogen ($f_{\HI}$), neutral 
helium ($f_{\HeI}$), singly ionized helium ($f_{\HeII}$), 
and the temperature (expressed in
terms of the total energy density $E=(3/2)n_{tot}k_{b}T$) 
according to the following rate equations:
\begin{align}
&\frac{\mathrm{d}f_{\HI}}{\mathrm{d}t}=-f_{\HI}\Gamma_{\HI}- n_{\mathrm{e}} f_{\HI} \beta_{\HI} C + n_{\mathrm{e}}f_{\HII}\alpha_{\HI} C\label{fHIeq}         \\
&\frac{\mathrm{d}f_{\HeI}}{\mathrm{d}t}=-f_{\HeI}\Gamma_{\HeI}- n_{\mathrm{e}} f_{\HeI} \beta_{\HeI} C + n_{\mathrm{e}} f_{\HeII}(\alpha_{\HeI}+\xi)C       \\
&\frac{\mathrm{d}f_{\HeII}}{\mathrm{d}t}=-f_{\HeII}\Gamma_{\HeII}- n_{\mathrm{e}} f_{\HeII} \beta_{\HeII} C + n_{\mathrm{e}} f_{\HeIII}\alpha_{\HeII} C \\
&\frac{\mathrm{d}E}{\mathrm{d}t}+2\mathrm{H}E = \sum_{i}^{3} n_i f_i G_i -\Lambda(n_i,T) \ .
\end{align} 
Here $\alpha_i$, $\beta_i$ and $\xi$ are the temperature-dependent 
radiative recombination (to all levels), collisional 
ionization, and dielectronic recombination coefficients, respectively;
$n_{\mathrm{e}}=n_{\mrm{H}}(1-f_{\HI})+n_{\mrm{He}}(2-2f_{\HeI}-f_{\HeII})$ is the number density of electrons, $C=\langle n^2\rangle/\langle n\rangle^2$ is 
the gas clumping factor\begin{footnote}{For simplicity, we assume the same value of $C$ for each species in the current version of the code.}
\end{footnote}, and $\Lambda$ is the total cooling rate (including
recombinations, collisional excitations, Compton, brehmstrahlung and Hubble cooling).
We use the analytical fits of Hui \& Gnedin (1998) for the ionization, recombination and cooling rates. 

One of the main challenges in the calculation of the photoionization/heating rate
is represented by the presence of the diffuse term in eq. (\ref{Isplit}), since it requires the transport
of the diffuse photons generated from atomic recombinations in every point of the medium. A widely used solution
to this problem is the so called ``on-the-spot'' approximation (OTS or \emph{Case B}, originally proposed by Baker \& Menzel 1938),  
which assumes that every diffuse, ionizing photon is absorbed exactly in the same point where it is generated (i.e., the photon mean-free-path is much smaller
than the resolution scale).
In this case, assuming for the moment a pure hydrogen medium, the diffuse part of the photoionization rate can be written as:
\begin{equation}
\Gamma_{\HI}^{\mrm{diff}}=n_e f_{\HI}\alpha^{1}_{\HI}C=n_e f_{\HI}(\alpha_{\HI}-\alpha^{\mrm{B}}_{\HI})C \ ,
\end{equation}
 where $\alpha^{1}_{\HI}$ is the recombination coefficient to the
hydrogen ground level, and $\alpha^{\mrm{B}}_{\HI}$ is the analogous coefficient for the levels $l>1$. 
Substituting this relation in eq. (\ref{fHIeq}), we obtain:
\begin{equation}
\frac{\mathrm{d}f_{\HI}}{\mathrm{d}t}=-f_{\HI}\Gamma^{\mrm{dir}}_{\HI}- n_{\mathrm{e}} f_{\HI} \beta_{\HI} C + n_{\mathrm{e}}f_{\HII}\alpha^{\mrm{B}}_{\HI} C \ ,
\end{equation}
i.e., a relation that depends only on the photoionization rate from individual sources ($\Gamma^{\mrm{dir}}_{\HI}$) and, thus, greatly simplifies the
radiative transfer problem. Analogous relations can be obtained for HeI and HeII, assuming that every ionizing photon from recombinations is absorbed
by the same species from which it originates (obviously, this approximation breaks down if we consider that non-self-ionizing photons from HeII recombinations
can actually ionize both HeI and HI, see section \ref{RecRadSec2} for a detailed discussion). 

 As also noticed by Ritzerveld (2005), despite being widely used, the OTS
approximation is based on an incorrect argument: if the local mean-free-path is very small as assumed, i.e., the local optical depth
is very high for the diffuse photons generated locally, even higher the optical depth would be for the radiation coming from discrete, non-local sources 
(unless their spectral energy distribution is much harder). For instance, 
in the classical Str\"omgren sphere situation with a monochromatic source (see Test 1 below), 
the regime where the OTS approximation may hold corresponds to regions where the directional flux from the central source is not able to penetrate, i.e.
only at the extreme edge of the Str\"omgren sphere. 
 The opposite case (called \emph{Case A}) 
is represented by the situation in which the mean-free-path of the diffuse photons goes to infinity and, effectively, $\Gamma_{\HI}^{\mrm{diff}}=0$.
Despite the loss of photon conservation for a bounded region like a Str\"omgren sphere, this approximation is actually more correct than
the OTS in most situations as we will show later.

\section{Computational method}

 The computational volume in \cname is discretized in a (series of) regular, block-structured grid(s) where the 
physical properties of the medium are associated with a zone-centered grid element, i.e. with a \emph{cell}.
The current possible choices for the computational domain in \cname are: 
i) single, uniform Cartesian grid, ii) \emph{static} multi-mesh structure, 
iii) \emph{evolving} multi-mesh structure.
 In the \emph{static} case, the grid is spatially fixed to the
initial, single or multi-mesh structure obtained, e.g., from the output of an Adaptive Mesh Refinement (AMR) hydro-simulation. 
In the \emph{evolving} case, this initial structure is adaptively refined (unrefined) each time the cells 
satisfy a chosen refinement (unrefinement) criterion. The refining procedure is described in details in section \ref{AMRsec}.
The radiation field (from individual sources and from diffuse radiation) 
is discretized in a series of rays that are propagated through this single or multi-mesh structure using a ray-tracing
algorithm. Before discussing the ray-tracing method, it is necessary to describe more in detail the multi-mesh implementation in \cnamens.

\subsection{Multi-mesh in \cname}

Multi-mesh domains in \cname are composed of a nested hierarchy of rectangular grids of different sizes and levels of refinement,
following the implementation called ``patch-based AMR'', originally described in Berger \& Oliger (1984).
The position and aspect ratios of this patches are optimized in order to increase the spatial (and temporal) resolution where needed
within the computational box.
Other possible multi-mesh implementations have been proposed in the past, like, e.g, the ``tree-based AMR'' (see Khokhlov 1998 and
references therein). In this case,
each cell (or $n^3$ group of cells in three-dimensions; typically $n=2$) is refined into children cells, on a cell-by-cell basis, and a ``grid'' 
is the analogous of a single cell (or a $n^3$ group of cells). These methods were originally developed for
the numerical solution of hydrodynamical equation and they have advantages and disadvantages with respect to each other for this
particular problem. In our case, patch-based AMR presents a significant advantage with respect to other implementations: it reduces
the total number of grids. One of the major problem presented by a ray-tracing algorithm is the fact that
a significant part of the computational time may be spent in ``crossing'' the simulation box. This is in a sense unavoidable, because 
solving the radiative transfer equations is a highly non-local problem. Optimizing the data and memory structure is thus fundamental in order to 
have an efficient algorithm. Minimizing the number of ``crossing'' grids helps reducing the overhead associated with the ray-tracing
in a multi-mesh structure. 

In order to increase the efficiency in ``crossing'' the computational volume, 
grids and cells in \cname at different levels of refinement are connected with two separate hierarchical trees and linked lists. 
Grids are strictly nested, i.e., each grid at level $l$ is fully contained by a single grid at
level $l-1$ (the ``parent'' grid). Note that, although this may increase the total number of grids, it allows a more efficient 
tree-search for both grids and
cells. Grids that share the same ``parent'' are connected via a linked list. Thus, each grid $g_i$ at level $l$ may have a maximum 
of three associated pointers: i) the ``parent'' grid at level $l-1$, 
ii) the ``next brother'' grid in the linked list at level $l$, iii) the ``son'' grid at level $l+1$
(i.e., the head of the linked list of grids fully contained within $g_i$). 
Grids without an associated ``son'' are called ``leaf'' grids. 
This hierarchical tree is a light structure that requires small amounts of memory
(if the number of grids is significantly smaller than the total number of cells, as it is always the case for ``patch-based AMR'') and that can be
reconstructed or saved easily given its strict nesting. If refined, a cell at level $l$ contains a group of $r_l^3$ subcells, where $r_l$ 
is the refinement factor. In principle, $r_l$ may vary with levels, although commonly is fixed to a constant value (typically $r_l=2$).
Analogously to the grid case, a cell without refinement is called a ``leaf'' cell.

Combined with the grid-tree, there is another simple hierarchical structure in \cname associated directly with the cells: 
each refined cell at level $l$ points to the grid at $l+1$ that contains its $r_l^3$ subcells. This simple (but more memory-consuming) 
cell-tree is built efficiently when needed from the grid-tree and, for this reason, is not saved during output or back-up in order
to save memory usage.  
The use of the cell-tree allows to recover
very quickly the location of the relevant subgrid (and thus the subcell) at level $l+1$ when crossing the tree towards higher levels of refinement,
without searching through the ``next brother'' linked list of the subgrids. When searching for cells at coarser (or at the same) level, the grid-tree
is used instead.
The combination of the grid-tree and the simple cell-tree represents a good balance between flexibility (that translate into computational speed) 
of the box crossing algorithm and memory consumption of the tree-structure.

\subsection{Ray-tracing algorithm}

 Ray-tracing is a well-studied problem in computational geometry, with several efficient methods commonly used in computer graphics. One of the
most efficient solution is the simple and fast grid-traversal algorithm of Amanatides \& Woo (1987), particular suited in the case of a single, uniform
Cartesian grid. Basically, this simple algorithm determines the intersection points between the boundaries of the cells and a ray 
(defined from its initial position and two direction angles) as it traverses the computational grid.

 We employ the grid and cell hierarchical trees discussed above to extend
this fast algorithm from the single, uniform grid to the multi-mesh structure. This is done simply adding to the original algorithm,
after each cell boundary crossing, a (recursive) check for the need of changing the current traversing grid (either at the same or at a different level). 
Let us consider, for example, the case
in which a ray is currently traversing a grid at level $l=l_i$.  
There are two situations in which a change of grid is required: 
a) the ray is still within the boundaries of the current grid but is entering a subgrid at level $l>l_i$;
b) the ray is exiting the boundaries of the current grid. 
We know that we are in the first situation
simply checking the current cell tree: if the cell is associated with a subgrid, we immediately recover from the tree the memory
location of the new grid at $l=l_i+1$ and, from the intersection coordinates, 
the corresponding cell within this grid. We recursively continue the search until we reach a ``leaf'' cell.
In the second case, the search consists of two phases. First, we cross the grid tree towards the ``parent'' grids at level $l<l_i$
until we find a grid where the ray is fully contained (and not at the grid boundaries) or 
where the ray is \emph{entering}\footnote{this is simply determined from the ray direction.} the grid boundaries. Note that usually
this search is very fast given that, for a typical nested structure, we only need to go to the level $l=l_i-1$.
When the grid at $l<l_i$ is found, we check if the cell where the ray is currently located is a ``leaf''. If this is not
the case, we apply the same procedure as in case a).

 When traversing a ``leaf'' grid we can apply the original traversal algorithm without the need for the recursive check (and change of grids) 
described above. In this respect, the use of a ``patch-based AMR'', given its larger grids, result in a reduction of the computational costs
of the traversing algorithm.

\subsection{\cname step by step}

 \cname is based on a ray-tracing algorithm that propagates the
(discretized) radiation field through the computational volume.
However, in the same spirit of the AMR method, the key element in \cname is not represented
by the \emph{ray} or photon package  
but, instead, by the computational \emph{cell}. In other words, the propagation of the radiation field 
(and the solution of the radiative transfer equation) is performed on a ``cell-by-cell''
basis rather than ``ray-by-ray''. Computationally, this translates into the change of the main loop
in the algorithm, from rays to cells: the radiation field from the (discrete and diffuse) sources is propagated with a photon-conserving
 method separately for each cell in the computational volume. As we explain below, at the heart of the
algorithm there is a new method that allows to draw a series of rays from a source, located at any position
inside (or outside) the box, through a cubic cell with the
\emph{correct} solid angle distribution. This extends the ``Monte Carlo'' methods, where a series of rays is uniformly
casted around a source to obtain the correct solid angle distribution, from the computational box to the cell level.

 This ``cell-by-cell Monte Carlo'' is a more natural approach when most of the physical evolution
of the medium happens rapidly only in a small portion of the simulated box at a time. This is a 
typical case in cosmological simulation, either because of a difference in density 
(e.g., self-shielded dense clumps in a mostly ionized medium) or in the ionized state 
(e.g., the expansion of an Ionization-front in a mostly neutral IGM during reionization) 
With the cell-based approach, we can split the computational volume, either single or multi-mesh, grouping the cells with different 
evolving times or properties and focusing the computational efforts only where actually needed.
This is obtained associating to each cell an individual time-step ($\Delta t_{c}$) 
that is used to determine whether the cell can be considered \emph{active}
or not during the current simulation step. We discuss how we derive $\Delta t_{c}$ and how we define the \emph{active} cell
in section \ref{TimeStepSec}.

In detail, the computational algorithm consists of an iterative method divided into four main parts for each
simulation step (and for each \emph{active} cell): 
i) finding the ionization and photo-heating rates, ii) choosing the time-step,
iii) solving the chemistry equation for the evolution of the
medium properties, and, if needed, iv) performing an adaptive refinement of the computational mesh.
 The first part of the algorithm is the most time-consuming (and
important) part of the method and is described in detail in the following section.

\subsubsection{Finding the ionization and photo-heating rates}\label{gammasec}

The photo-ionization ($\Gamma_{i}$) and photo-heating rates ($G_i$) are computed for each 
\emph{active}, leaf cell (during the current time-step)
with an iterative Monte Carlo procedure until convergence for both quantities is reached. 
The convergence level is typically set at 1\%, but can be increased (decreased) depending
on the requested level of accuracy.

We compute separately, with two different methods, 
the contribution to the total photoionization/heating rate from discrete sources ($\Gamma^{\mrm{dir}}_i$)
and from diffuse emission ($\Gamma^{\mrm{diff}}_i$). Most of the effort is dedicated 
to the calculation of ($\Gamma^{\mrm{dir}}_i$), since in many (cosmological) cases
the directional flux from discrete sources represents the major contribution to the total photoionization rate.

The procedure to obtain the values of $\Gamma^{\mrm{dir}}_i$ and $G^{\mrm{dir}}_i$ is the following:
for each
iteration step, a packet of rays is propagated from selected points within the cell to
the sources. These points are chosen with a Monte Carlo procedure
based on a rejection algorithm 
that insures an uniform solid-angle distribution within the cell. The total flux is then
re-scaled according to the fraction of the solid angle covered by the cell. We present in the Appendix the
analytical approximation that gives the solid angle of a (cubic) cell
as seen by any point inside (or outside) the box.

 For each ray, we compute the HI, HeI, and HeII column densities both 
for the path length within the cell ($\Delta N_i$) and for the path
length connecting the cell edge to the source ($N_i$).
 The column densities are then used to calculate the frequency-dependent
optical depths $\Delta \tau_i(\nu)=\sigma_i (\nu) \Delta N_i$ 
and $\tau_i(\nu)=\sigma_i (\nu) N_i$.  
Given the optical depths, we derive the probability that a photon with frequency $\nu$ 
is absorbed by the species $i$ within the cell:
\begin{equation}
P_i(\nu)=
\frac{1-\mathrm{exp}[-\sum\limits_{j=1}^{3}\Delta \tau_j(\nu)]}
{\sum\limits_{j=1}^{3}\{1-\mathrm{exp}[-\Delta\tau_j(\nu)]\}}
\{1-\mathrm{exp}[-\Delta\tau_i(\nu)]\}\ .
\end{equation}
The optical depths and the 
spectral energy distribution of 
the sources are sampled into $N_{\nu}$ (logarithmically-spaced) 
frequency bins. 
The probability distributions $P_i$ are used to compute the photoionization rate,
including for the moment only the first term in eq. (\ref{Ieq}),
for each ray of the packet and for each source:
\begin{equation}
\tilde{\Gamma}^{\mrm{dir}}_i=\sum\limits_{i_{\nu}=1}^{N_{\nu}}
P_i(\nu_{i_{\nu}})
\dot{N}_{\gamma}(\nu_{i_{\nu}})
\mathrm{exp}[-\tau_i(\nu_{i_{\nu}})]\ ,
\end{equation}
where $\dot{N}_{\gamma}(\nu_{i_{\nu}})$ denotes the number of ionizing photons
per unit time emitted by the source in the frequency bin $i_{\nu}$. Finally,
the value of $\Gamma^{\mrm{dir}}_i$ is obtained by averaging $\tilde{\Gamma}^{\mrm{dir}}_i$ over the number
of rays in the packet, the cell volume and the fraction of the solid angle
covered by the cell (see Appendix).
Several packets of rays are generated until the value
of $\Gamma^{\mrm{dir}}_i$ converges to the required level. The procedure is repeated for each source and
the single values of $\Gamma^{\mrm{dir}}_i$ are added. 
The photoionization rates $G^{\mrm{dir}}_i$ are obtained in a
similar way, starting from 
\begin{equation}
\tilde{G}^{\mrm{dir}}_i=\sum\limits_{i_{\nu}=1}^{N_{\nu}}
P_i(\nu_{i_{\nu}}) h(\nu_{i_{\nu}}-\nu_{i})
\dot{N}_{\gamma}(\nu_{i_{\nu}})
\mathrm{exp}[-\tau_i(\nu_{i_{\nu}})]\ .
\end{equation}

In order to avoid redundant calculations of the column densities $N_i$ from
the cell edges to the sources for each ray packet, we first evaluate $N_i$
on the cell vertexes ``visible'' from each source. If the difference
between these values is less than a given (small) threshold, we do not
use the full ray-casting algorithm. Instead, we use the $N_i$ values
at the vertexes to interpolate (linearly) the needed values on the cell faces.  
Typically, the cells that need the full ray-casting algorithm are a few percent
of the total volume and correspond to the Ionization-front regions (where
the column-density varies rapidly). For the remaining cells, interpolating
the column-densities is a good approximation and allows to substantially
speed-up the computation. Note that in this case, the algorithm is similar to 
a classical long-characteristic ray-tracing.

To achieve a better performance it is
also possible to choose a column-density threshold (or, equivalently, an optical
depth threshold at the ionization limit) above which \cname skips the calculation
of the $\Gamma^{\mrm{dir}}_i$ and $G^{\mrm{dir}}_i$ values. This is computationally convenient, e.g., in the
early stages of the expansion of an Ionization-front in a neutral and dense
medium, when most of the volume is still optically thick to the source radiation.
Analogously, it is possible to choose a minimum value of $\Delta N_i$ and $N_i$ 
below which an optically-thin approximation is used instead of the full ray-tracing
algorithm. This is typically the situation for the ionized region in proximity
of the sources, where the optical depth is negligible.

As mentioned in section 2, 
eq. (\ref{Ieq}) permits I-fronts to
propagate faster than light in near proximity of a bright source.
An exact solution for this problem that does not require to fully solve the time-dependent radiative transfer equation
and avoids the loss of photon conservation
is only available in the case of an homogeneous medium (e.g., Shapiro et al. 2006) as we have also discussed elsewhere (Cantalupo et al. 2008).
In the more general case of an inhomogeneous medium, we fix this unphysical behaviour 
by disregarding the ionizing flux of a given source at a distance $r>(t_s\cdot c)$, where $t_s$
is the source current lifetime. 
Note that, the loss of photon conservation due to this approximation is typically restricted to the very early phases
of the expansion of the I-front produced by very bright sources.

Once the value of the directional photoionization/heating rate have been obtained, we calculate (with a simpler
procedure) the diffuse component rates for the same, \emph{active} cells.
Also in this case we follow a ``cell-by-cell'' approach: given a particular (\emph{active}) cell
for which we want to compute $\Gamma^{\mrm{diff}}$ and  $G^{\mrm{diff}}$,
we generate with a Monte Carlo procedure a set of rays that propagates
outwards from the cell centre.
For each cell encountered by the
ray, we compute the medium opacity and the source term as a function of the $n_{e}$, $n_{\HII}$, $n_{\HeII}$, $n_{\HeIII}$, and temperature. 
In particular, we include: i) HI, HeI and HeII free-bound continuum (from Osterbrock 1989), 
ii) HeII Balmer continuum (from Ercolano \& Storey 2006), iii) HeII two-photon continuum (from
Nussbaumer \& Schmutz 1984), and iv) the HeII Ly$\alpha$ line.
We neglect the HeI Balmer continuum and HeI emission lines given their
(relatively) small contribution to the total emissivity. 
Given the source function, we compute and add the respective $\Gamma^{\mrm{diff}}$ and $G^{\mrm{diff}}$ to the direct
photoionization/heating rate, according to the equation presented in section 2.

 In this method, each ray represents a fixed fraction of the solid angle ($\theta_d$) of the sky as seen by the cell,
determined by the chosen number of rays (that we leave in the current version as a free parameter). Note that this procedure
is accurate if the three-dimensional 
diffuse field is homogeneous over a scale corresponding to $\theta_d\times r$, where $r$ is the distance from the cell.
Obviously, if the medium is clumpy, this method becomes more and more inaccurate at increasing distances
from the cell, unless the number of rays is increased accordingly. Note, however, 
that in most cosmological situations, the 
diffuse field represents a small component with respect to the direct radiation from sources.

\subsubsection{Choosing the time-step and the active cells}\label{TimeStepSec}

As mentioned above, 
we associate an individual time-step $\Delta t_{c}$ to each cell in the computational volume. 
We first estimate $\Delta t_{c}$ for the cells with the newly calculated $\Gamma_i$
by taking a fixed fraction $f$ of the minimum ionization timescale
from the rate equations. 
The minimum value found for $\Delta t_{c}$ is stored (as the new global time-step $\Delta t$). At this
point we check which cells can be considered \emph{active} during the next simulation
step comparing the individual $\Delta t_{c}$ with $\Delta t$. If $\Delta t_c<f_{ts}\Delta t$,
where $f_{ts}\geq1$ is a user-defined parameter, the cell is considered \emph{active} and
we assign $\Delta t_c=\Delta t$. If a previously \emph{active} cell does not meet this criterion we 
de-activate the cell until a simulation time $f_{ts}\Delta t$ (or a number $n_{ts}<f_{ts}$ of simulation steps) 
has elapsed. If $n_{ts}=1$ all (leaf) cells are always considered active.

 Since the cells within the Ionization-front have the shortest $\Delta t_{c}$, they are always
selected as \emph{active}. In practice, the factor $f_{ts}$ controls how broad is the region of \emph{active} cells
around the I-front. Cells outside of this region are updated only when necessary, typically after several simulation
time-steps, given their much longer ionization/recombination time-scales. Even for very large values of $f_{ts}$ (e.g., $f_{ts}=10^4$)
the gain in computational speed is huge, since most of the volume has evolution time-scales several orders of magnitude
larger than the I-fronts.

\subsubsection{Evolving the medium properties}

In this part of the computational algorithm, we solve the rate equation
and we evolve the medium properties for the \emph{active} cells with the newly calculated
$\Gamma_i$ and $G_i$. 
For the integration of the rate equations, 
we use the Radau IIA method (Hairer \& Wanner 1996), an implicit
Runge-Kutta scheme of variable, adaptive order. This allows this part of the code
to be computationally stable and reasonably fast for the required
accuracy.

\subsubsection{Adaptive refinement of the grid structure}\label{AMRsec}

If activated, the Adaptive Mesh Refinement is performed with the recursive clustering algorithm of
Berger \& Rigoutsos (1991). Before applying this algorithm, we flag the cells that satisfy a given refinement criterion.

The currently implemented refinement criterion is based on a combination of three variables for each atomic species:
i) the cell neutral fraction, ii) the cell ionization rate ($\Gamma_i$), and iii) the (frequency dependent) 
cell optical depth $\Delta \tau_c$.
The ultimate goal is to flag and refine the fast evolving cells that are inside the I-front (i.e., the cells
with the highest $\Gamma_i$), until their $\Delta \tau_{c}$ is below the desired value. In practice, the refinement
criterion should produce a multi-mesh grid where the front is well resolved, i.e. where the $\Delta \tau_c$ at the 
species energy threshold is well below unity. Moreover, we would like also to have a pre-refined region just beyond
the current position of the I-front, and a \emph{sharp} cut corresponding to the regions where the I-front
is just passed leaving the medium (highly) ionized. The size and shape of the refining region can be easily
tuned with a criterion that includes the three variables discussed above, see Test 6 for an example.

 The recursive AMR algorithm  starts from the base grid at the coarser level and it is composed by three main steps: i) flagging,
ii) clustering, and iii) refining/unrefining.  
 Once the cells to be refined at the coarser level ($l_i=0$) are flagged as discussed above, 
they are clustered into new rectangular patches. These are splitted until 
the clustering efficiency (i.e., the ratio between the number of flagged and total number of cell in the new patch)
is greater than a chosen minimum threshold, typically 80\%. In order to avoid the creation of grids that are too small (and thus inefficient
for the ray-tracing algorithm), it is also possible to chose a minimum size for the newly created patches. In most of the
situations, we find that a minimum size of two (parent) cells gives the best balance between number of newly created cells and efficiency
of the algorithm. We then \emph{propagate} the physical properties from the coarser grid ($l^i$) to the newly created patches 
at level $l^{i+1}$ (refinement). Currently, only straight injection refinement has been implemented. 
If a cell at level $l^i$ that was previously refined has not been flagged during the current time-step, the properties of its
subcells at level $l^{i+1}$ are \emph{propagated} back to the parent cell that becomes now a leaf cell (un-refinement).
Temperature during un-refinement is weighted according to the subcell electron number densities, 
while the species neutral fractions ($y_i$) are weighted according to the mass densities. 
 This algorithm is recursively repeated, patch by patch, at finer levels, until there are newly flagged cells or the maximum
number of refinement levels is reached. 

 During clustering and refinement, the grid-tree and the cell-tree are updated, new memory is dynamically allocated for the newly created
grids and, at the same time, the memory slots associated with the un-refined grids are released. By the algorithm construction,
parent and son grids have memory slots allocated in close parts of the memory map. This ensures an efficient memory usage and reduces
computational overhead during ray-tracing. 

 In case where the AMR is performed on an existing multi-mesh grid (e.g., the output of an hydrodynamical AMR code), there is the
very important option in \cname 
to consider this initial grid as \emph{fixed}: i.e.,  the AMR builds the new patches on the top of the existing multi-mesh
grid without unrefining it below the original, initial level. This 
allows us to increase the resolution where needed for the radiative transfer
(on the I-Front) without losing the resolution achieved with the hydro code.
The capacity of \cname of adding further levels of refinement on an existing AMR grid, without loosing previous information,
is of great importance for several physical applications. For example for the study of the highly ionized 
(but denser than the average) regions in proximity of a QSO that where already touched by its I-front 
and that would appear as Ly$\alpha$ forest in the quasar spectrum.

\begin{figure}
\begin{center}
\includegraphics[totalheight=0.27\textheight]{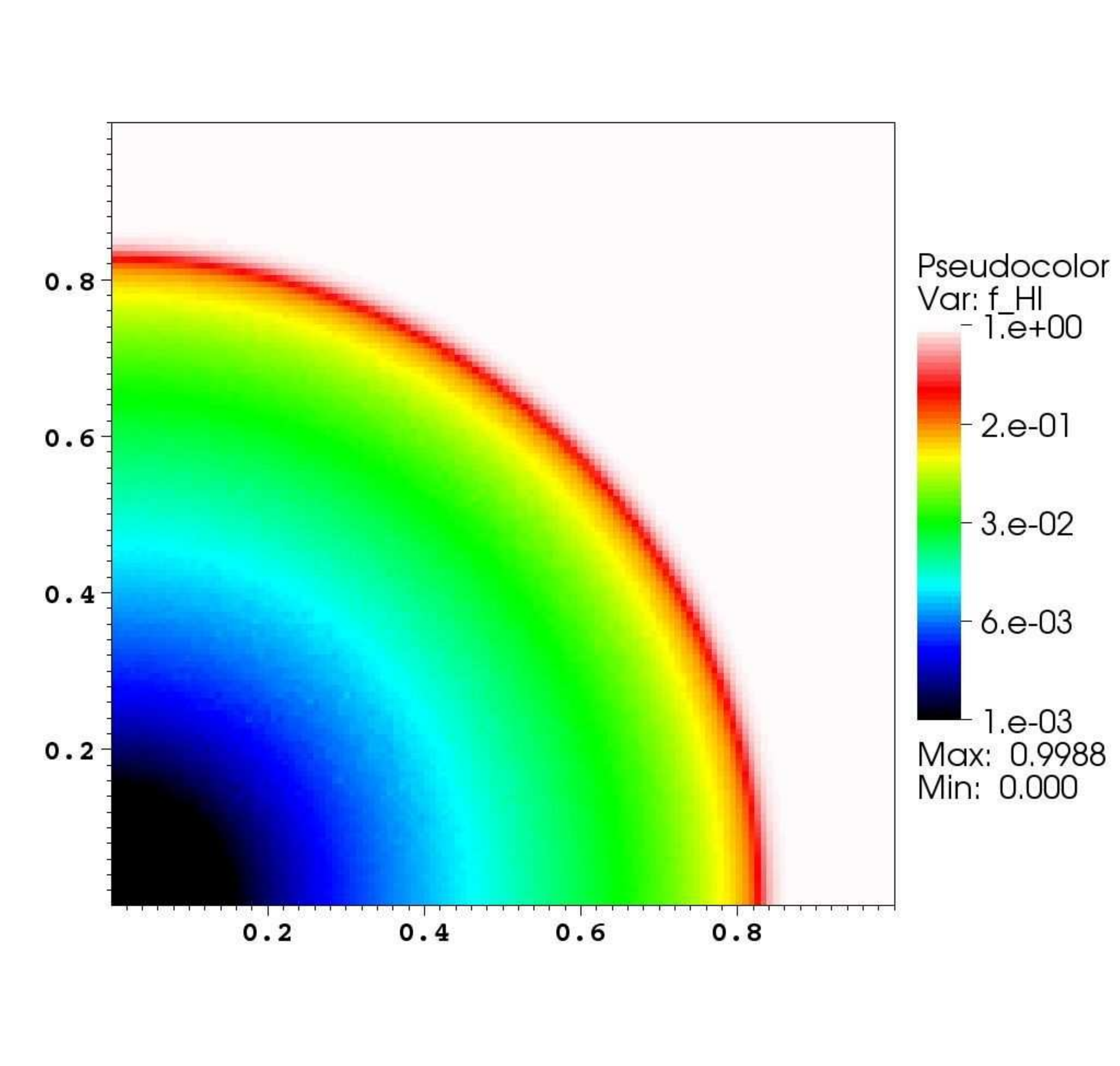}
\includegraphics[totalheight=0.27\textheight]{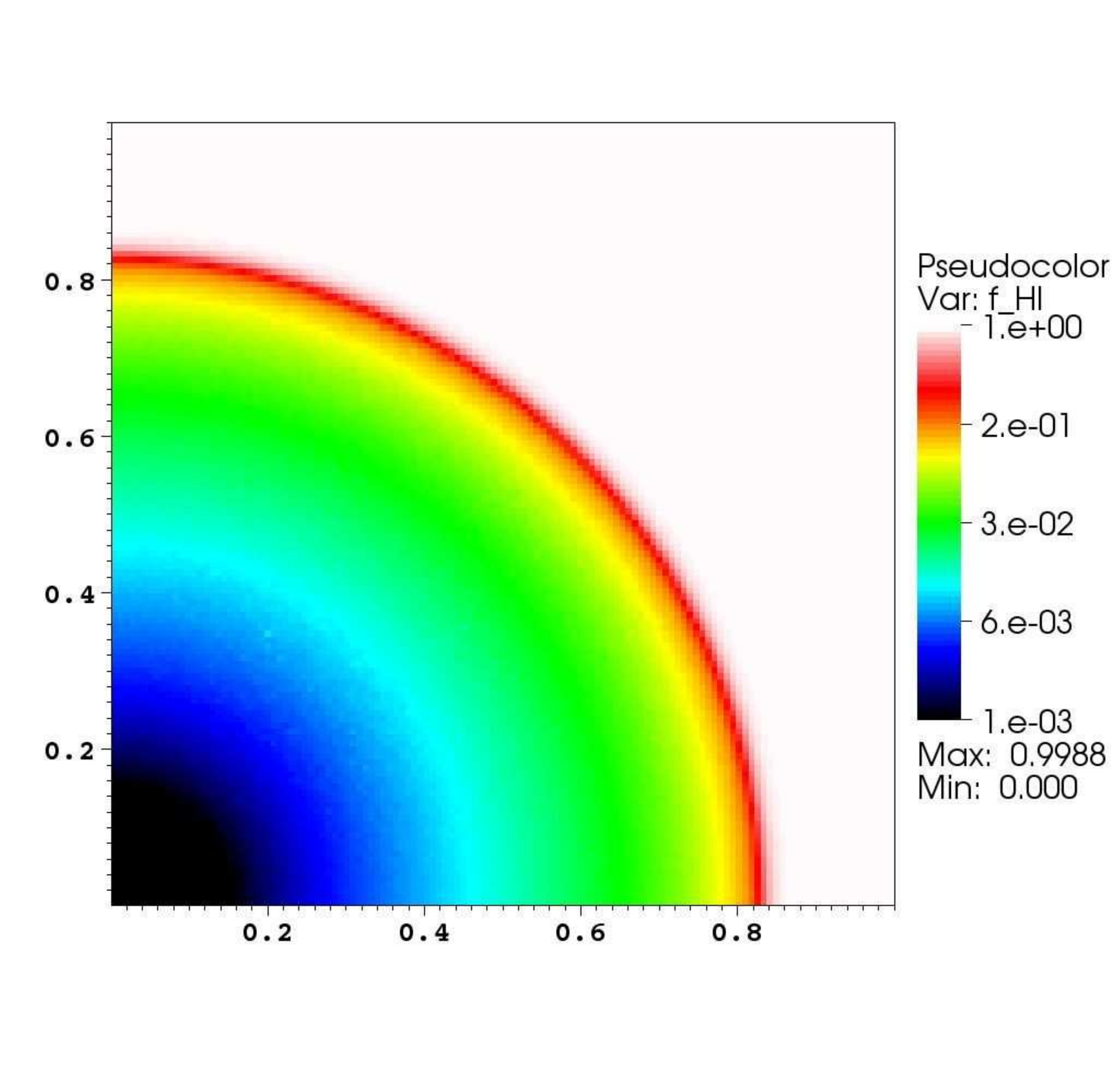}
\caption[Test 1: HI fraction images]
{Test 1: image slices of the HI fraction at time $t=500$ Myr in the plane
$y=0$ (top panel) and $z=0$ (bottom panel).}
\label{T1}
\end{center}
\end{figure}

\begin{figure}
\begin{center}
\includegraphics[totalheight=0.27\textheight]{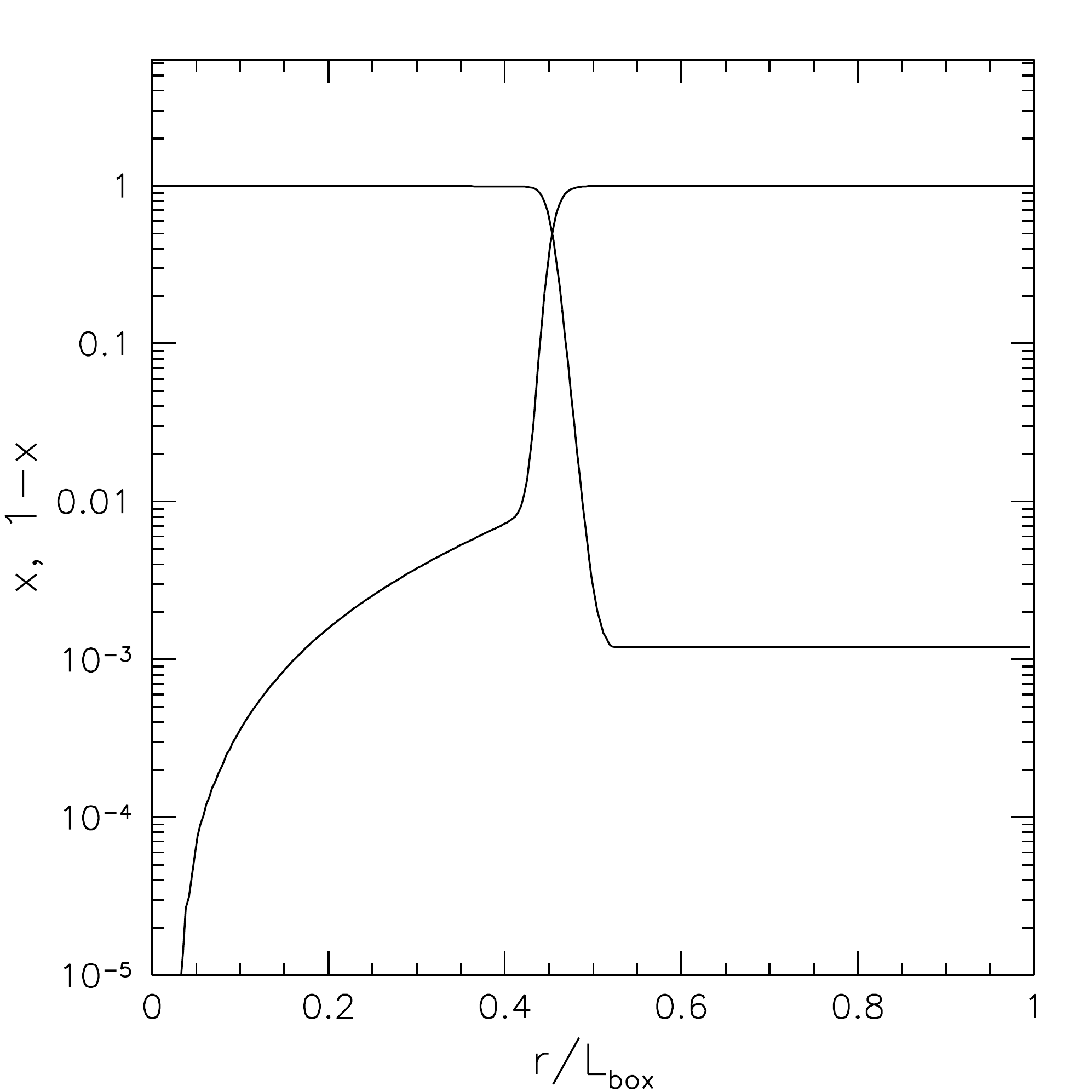}
\includegraphics[totalheight=0.27\textheight]{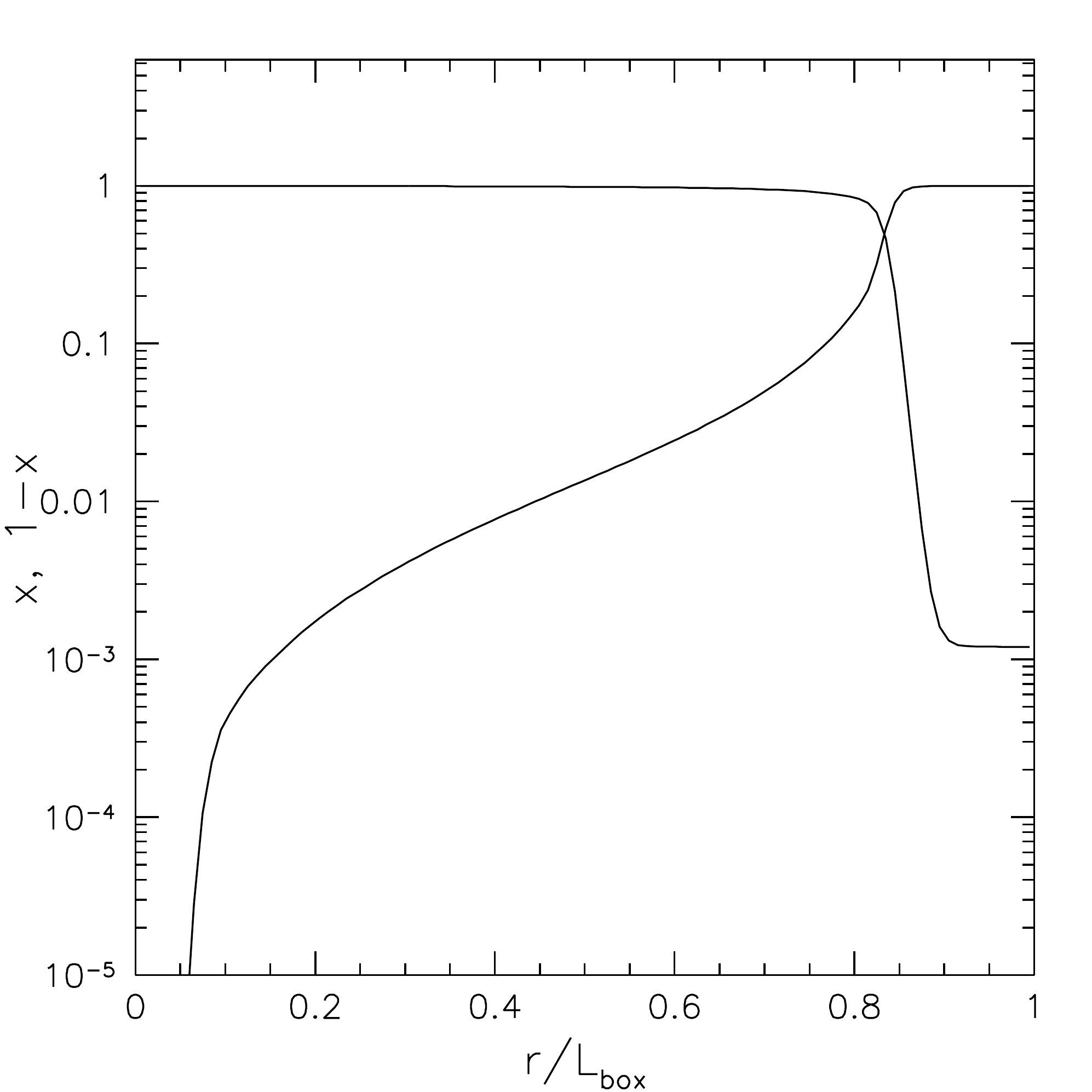}
\caption[Test 1: spherically averaged HI profiles]
{Test 1: spherically averaged ionized (x) and neutral (1-x) fraction profiles at times
$t=30$ Myr (top panel) and $t=500$ Myr (bottom panel).}
\label{T1f8}
\end{center}
\end{figure}
\begin{figure}
\begin{center}
\includegraphics[totalheight=0.27\textheight]{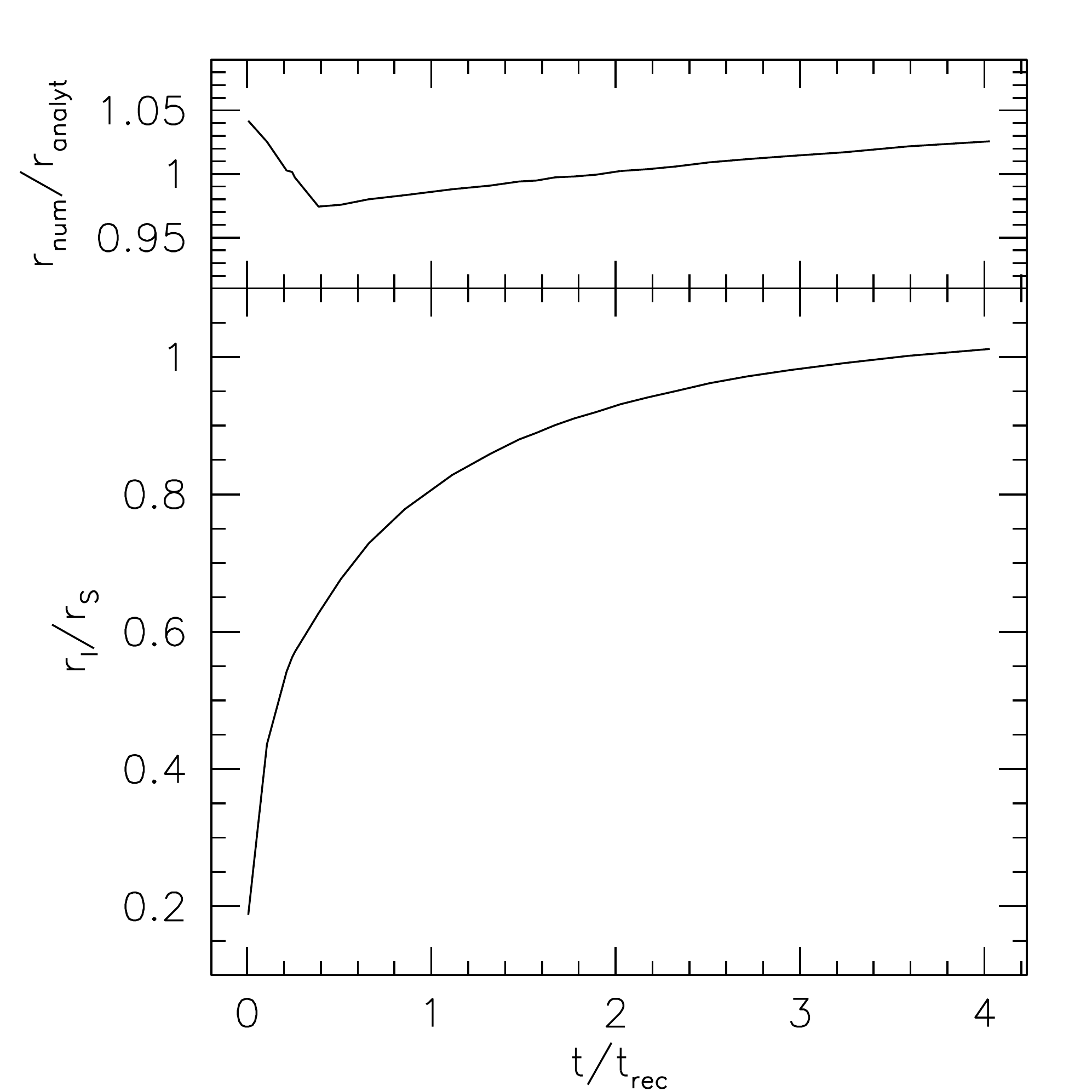}
\caption[Test 1: evolution of the I-front position]
{Test 1: time-evolution of the I-front position $r_I$ with respect
to the Str\"omgen radius $r_S$ (bottom panel) and the analytical solution (top
panel). See text for details.}
\label{T1f7}
\end{center}
\end{figure}

\section{Validating tests}

In this section we present the validating tests of the code based on the
radiative transfer code comparison project (Iliev et al. 2006; I06 thereafter). These
tests have been designed in order to compare all the
important aspects of several radiative-transfer codes present in the
literature. They include the correct tracking of both slow and fast Ionization-fronts
in homogeneous and inhomogeneous density fields, I-front trapping, spectrum hardening and
the solution of the temperature state. The original tests in I06 (Test 1 to 4)
are performed for a single, uniform grid, pure hydrogen medium and without recombination radiation (using the
OTS approximation). 
In order to compare our results with the other RT codes, 
we use the same single grid (pure-hydrogen) configuration as used in I06 to reproduce the results of Test 1 to 4.
In the second part of this section, we present a set of case studies aimed at substantiating the
new characteristics of our code. In particular, we verify in Test 5 and Test 6 the multi-mesh,
adaptive capability of \cnamens. In Test 7 and Test 8 we show the effects of the diffuse radiation transfer for
hydrogen only medium (Test 7) and including helium (Test 8).

\begin{figure}
\begin{center}
\includegraphics[totalheight=0.27\textheight]{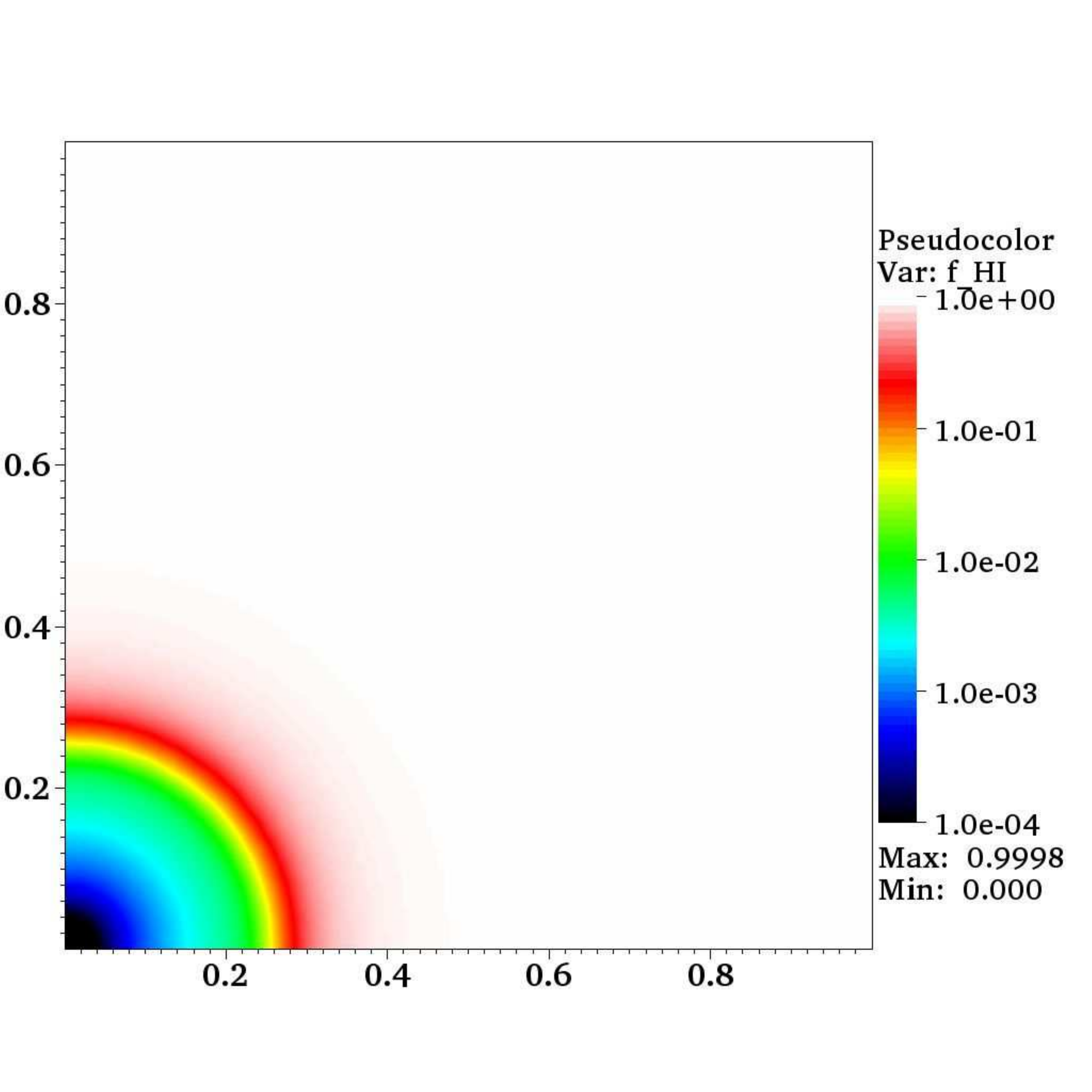}
\includegraphics[totalheight=0.27\textheight]{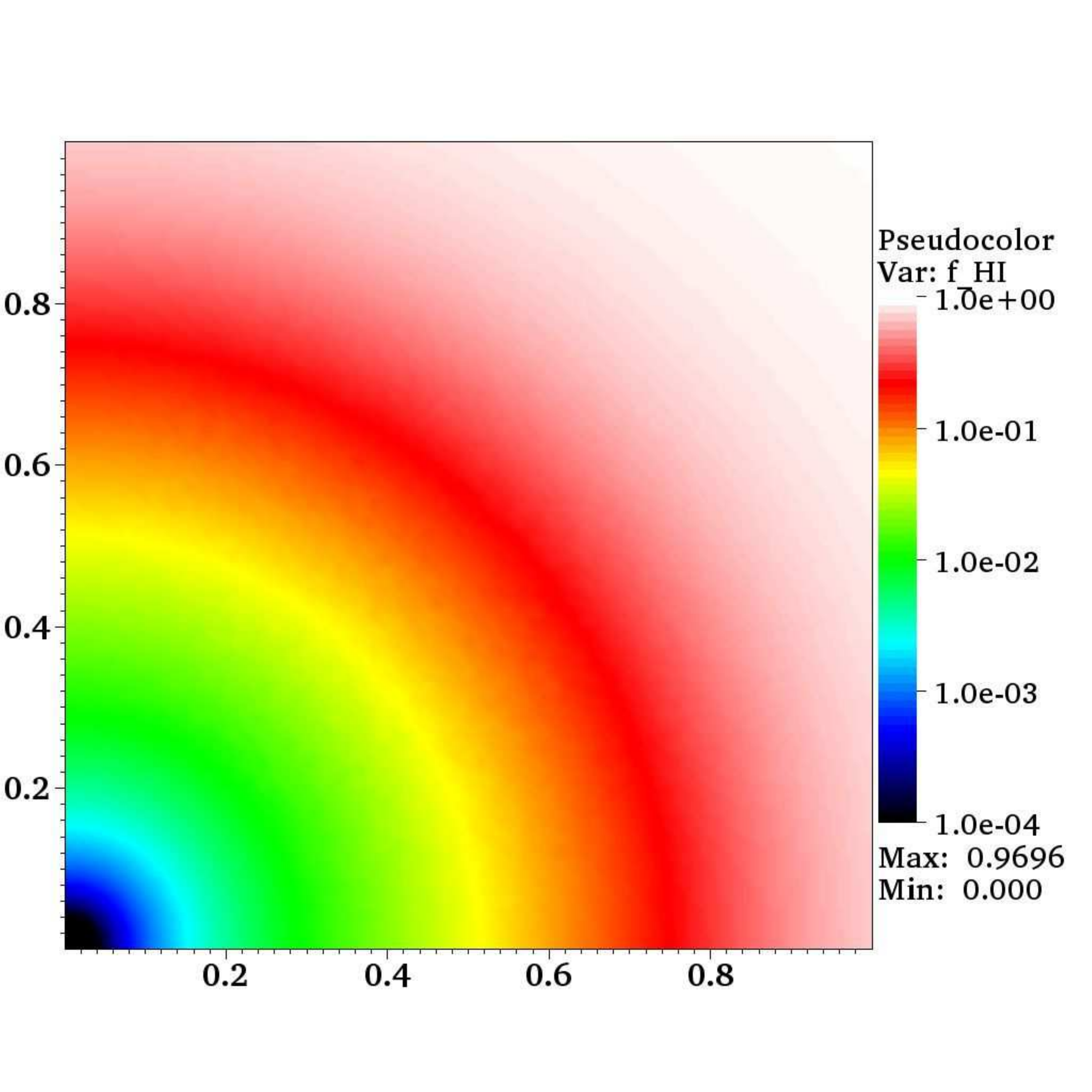}
\caption[Test 2: HI fraction images]
{Test 2: image slices of the HI fraction at coordinate $z=0$ and times
$t=10$ Myr (top panel) and $t=100$ Myr (bottom panel).}
\label{T2HI}
\end{center}
\end{figure}
\begin{figure}
\begin{center}
\includegraphics[totalheight=0.27\textheight]{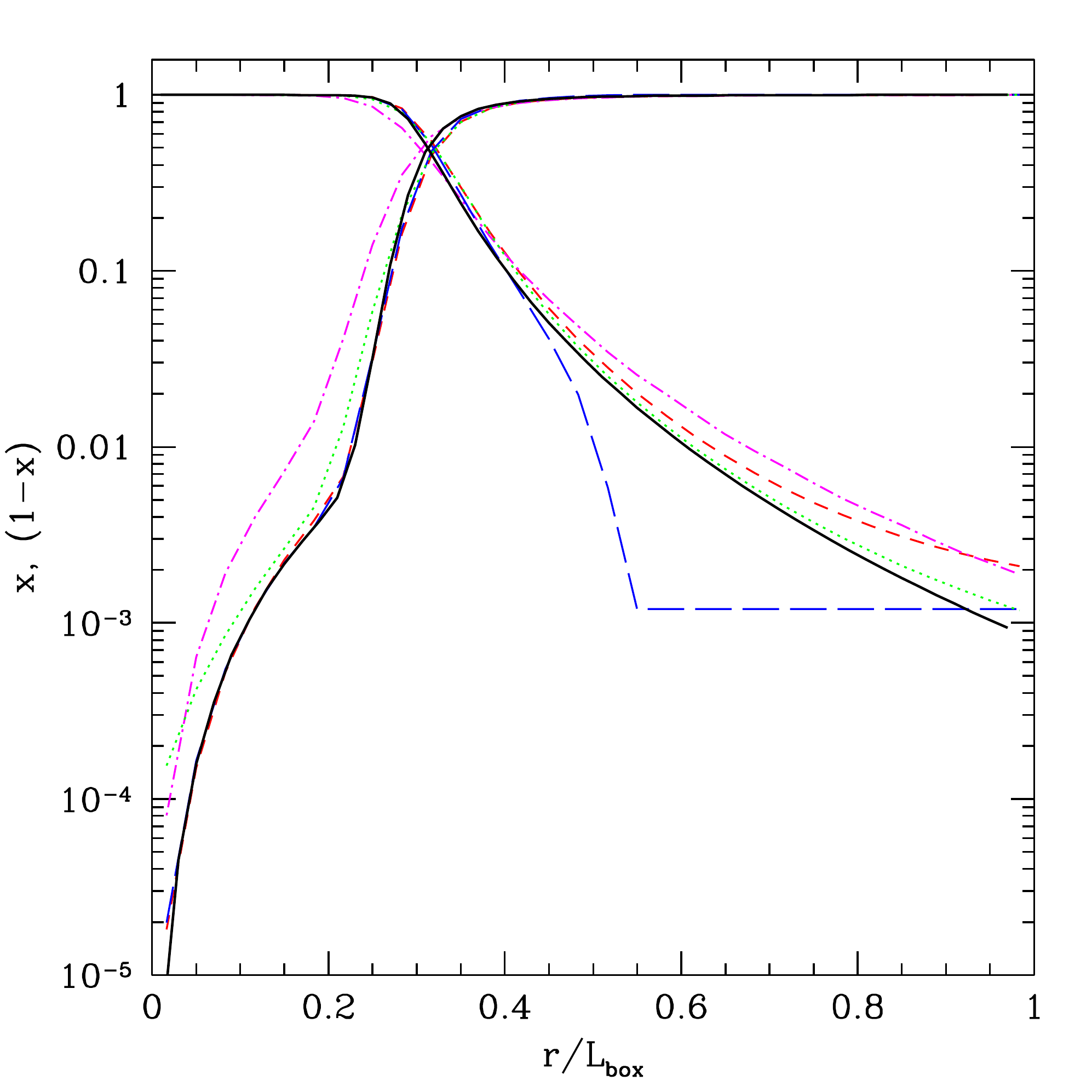}
\includegraphics[totalheight=0.27\textheight]{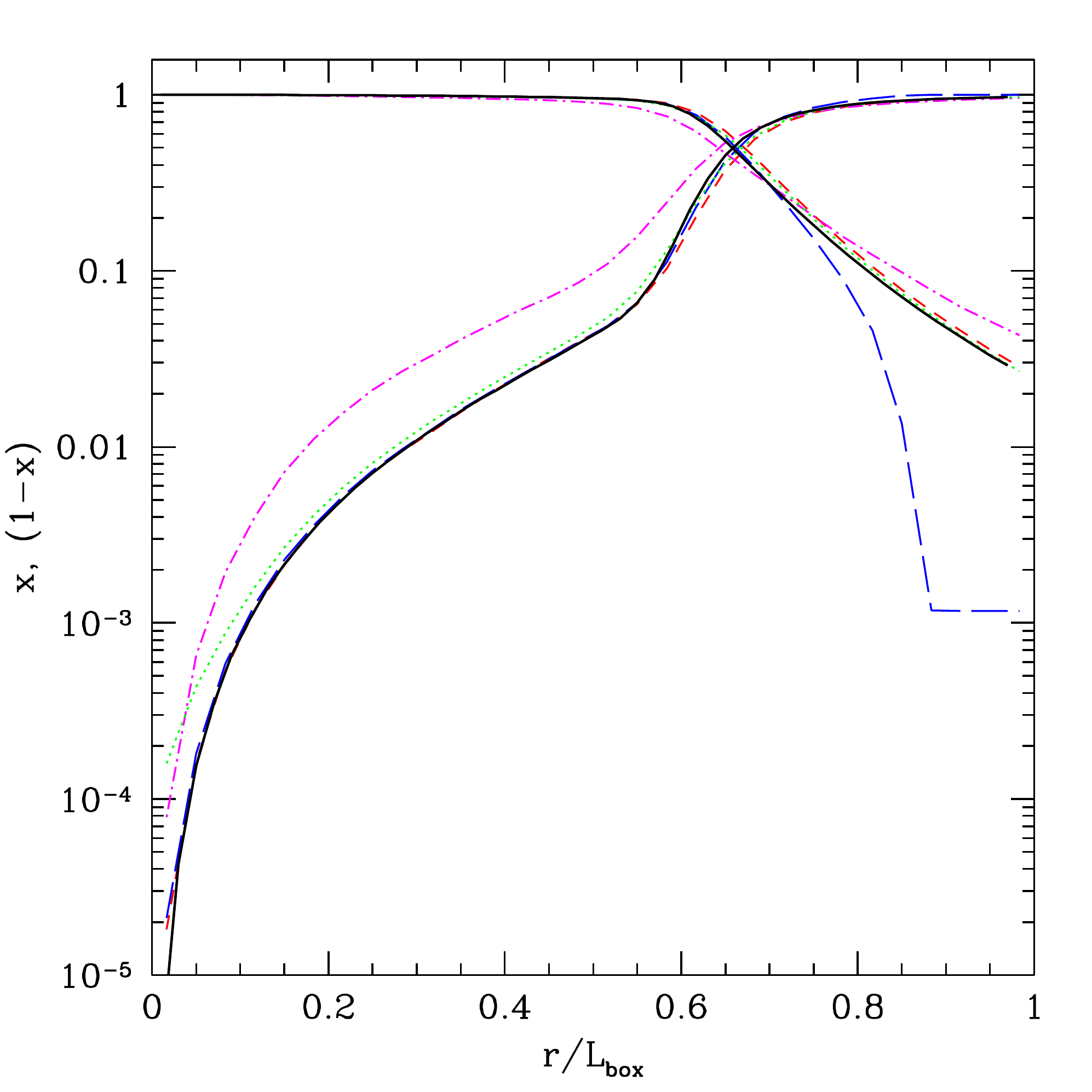}
\caption[Test 2: spherically averaged HI profiles]
{Test 2: spherically averaged ionized (x) and neutral (1-x) fraction profiles at times
$t=10$ Myr (top panel) and $t=100$ Myr (bottom panel).
The results from \cname are presented as solid black lines. The other
lines represent the results from a sample of four different codes 
taken from the RT code comparison project (I06), in particular: \texttt{C$^2$-RAY} (red,
short-dashed line), \texttt{CRASH} (cyan, dotted line), \texttt{FTTE}
(blue, long-dashed line), and \texttt{RSPH} (green, dot-dashed line). 
}
\label{T2f16}
\end{center}
\end{figure}

\begin{figure}
\begin{center}
{\includegraphics[totalheight=0.27\textheight]{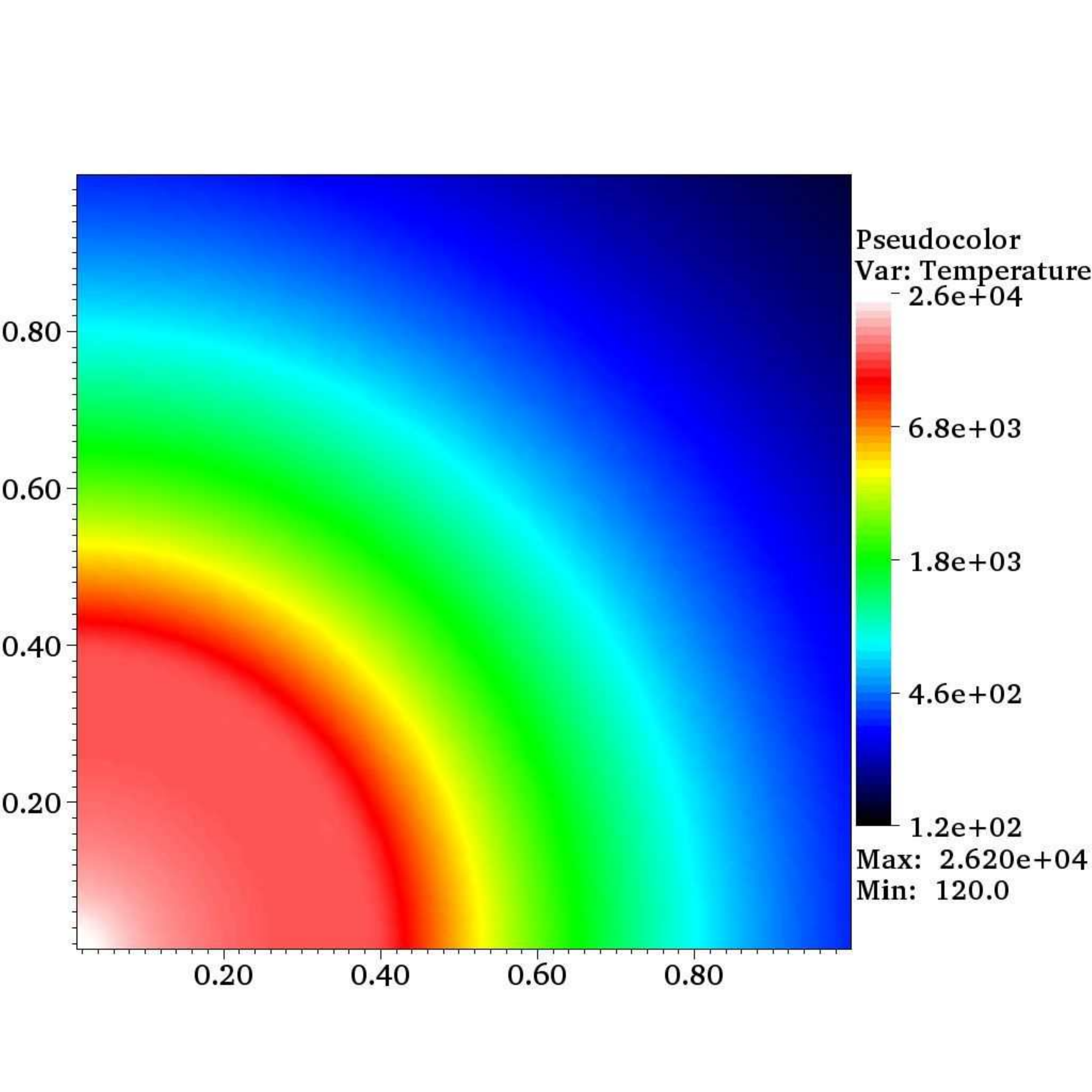}}
{\includegraphics[totalheight=0.27\textheight]{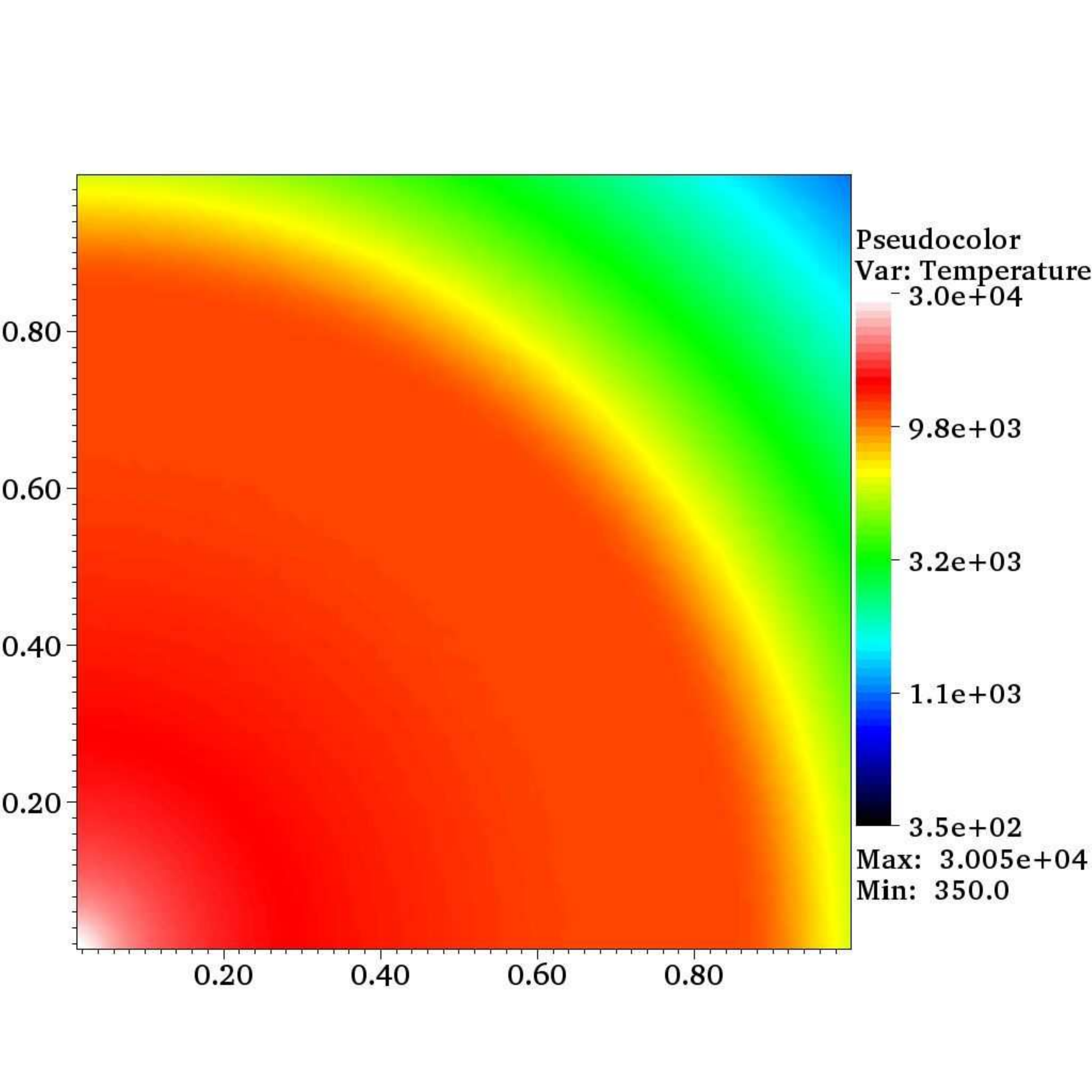}}
\caption[Test 2: Temperature images]
{Test 2: image slices of the gas temperature at coordinate $z=0$ and times
$t=10$ Myr (top panel) and $t=100$ Myr (bottom panel).}
\label{T2Temp}
\end{center}
\end{figure}
\begin{figure}
\begin{center}
{\includegraphics[totalheight=0.27\textheight]{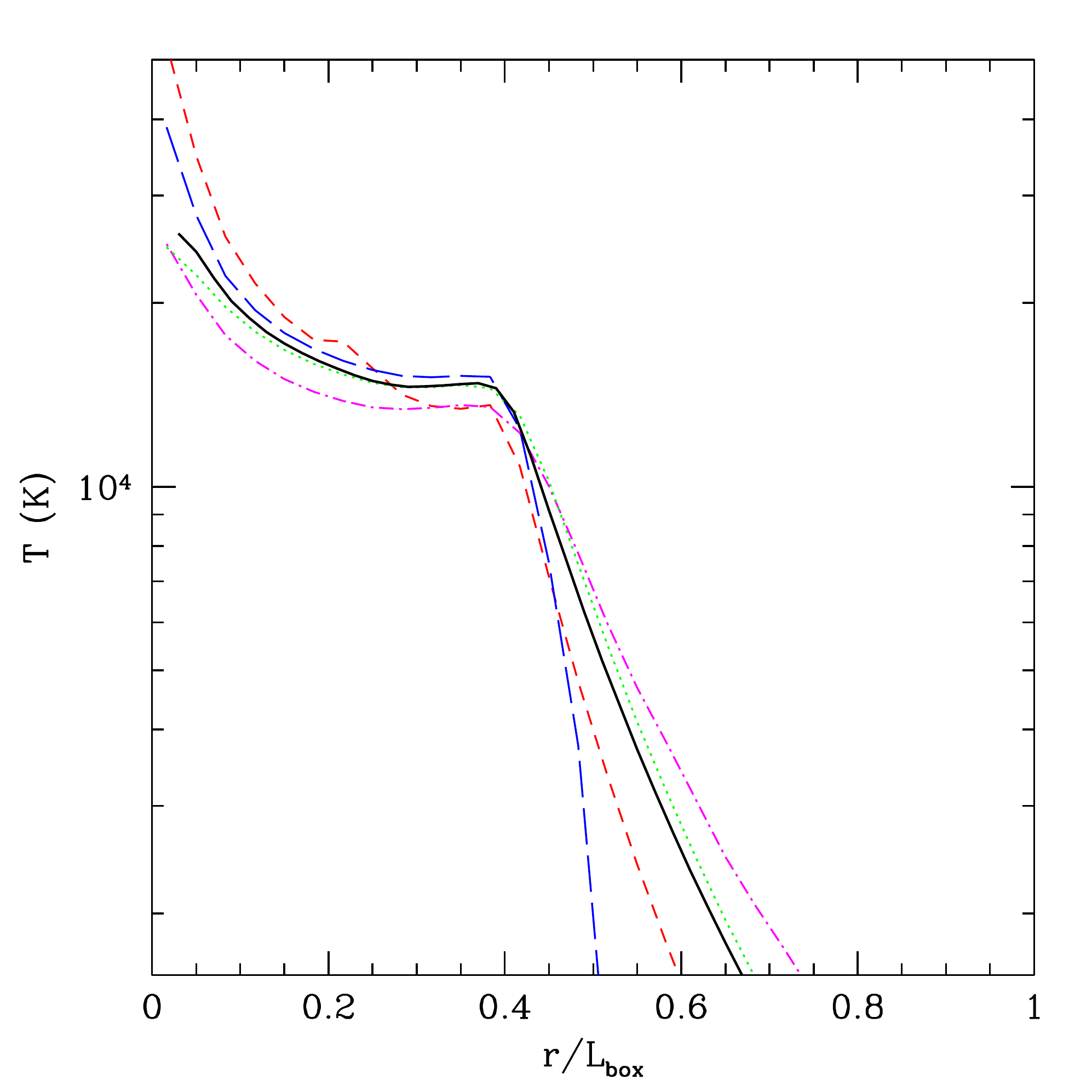}}
{\includegraphics[totalheight=0.27\textheight]{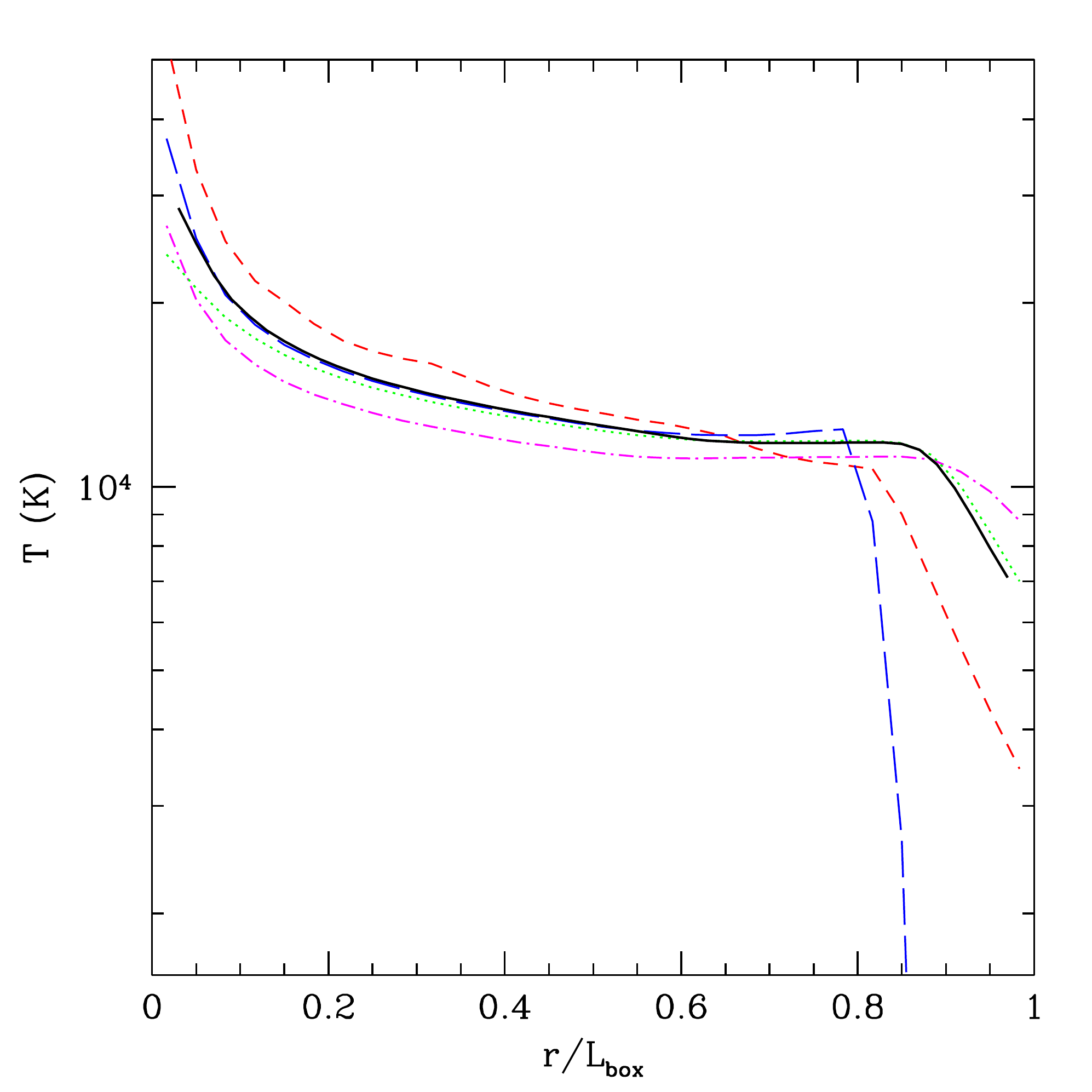}}
\caption[Test 2: spherically averaged temperature profiles]
{Test 2: spherically averaged temperature profiles at times
$t=10$ Myr (top panel) and $t=100$ Myr (bottom panel). 
The results from \cname are presented as solid black lines. The other
lines represent the results from a sample of four different codes 
taken from the RT code comparison project (I06), see caption in Figure \ref{T2f16}. 
}
\label{T2f17}
\end{center}
\end{figure}

\subsection{Test 1: isothermal HII region expansion}

This test represents the classical problem of the expansion of a HII
region in an uniform (pure-hydrogen) medium around a single ionizing source.
We assume that a steady, monochromatic
($h\nu=13.6$ eV) source emitting $\dot{N}_\gamma$ ionizing photons per unit
time turns on in an initially-neutral, uniform-density, static
medium with hydrogen number density $n_H$. Likewise in I06, we use for this
test the OTS approximation. 
The temperature is fixed at $T=10^4$ K. Under these conditions, and  
assuming that the front is sharp (i.e. that it is infinitely-thin, with the 
gas inside fully-ionized and the gas outside fully-neutral), there is a 
well-known analytical solution for the evolution of the I-front radius, 
$r_I$, and velocity, $\rm v_I$, given by 
\begin{eqnarray}
r_I&=&r_{\rm S}\left[1-\exp(-t/t_{\rm rec})\right]^{1/3}\,,\\
\rm v_I&=&\frac{r_{\rm S}}{3t_{\rm rec}}\frac{\exp{(-t/t_{\rm rec})}}
{\left[1-\exp(-t/t_{\rm rec})\right]^{2/3}}\,,
\label{strom0}
\end{eqnarray} 
where
\begin{equation}
r_{\rm S}=\left[{3\dot{N}_\gamma\over 4\pi \alpha_B(T) n_{\rm
  H}^2}\right]^{1/3}\,,
\end{equation}
is the Str\"omgren radius, i.e. the radius
at which recombinations balance the ionizations and the
HII region expansion stops.
 Here $\alpha_B(T)$ is the Case B recombination coefficient and 
\begin{equation}
t_{\rm rec}=\left[\alpha_B(T) n_{\rm H}\right]^{-1}\,,
\end{equation}
is the recombination time. The HII
region initially expands quickly and then slows down as the evolution 
time approaches the recombination time, $t\sim t_{\rm rec}$.
At a few recombination times, the I-front
stops and in absence of gas motions remains static
thereafter.

The numerical parameters for this test are the followings: 
computational box dimension $L=6.6$~kpc (the source is at one corner
of the box), gas number density 
$n_H=10^{-3}$~cm$^{-3}$, initial ionization fraction (given by collisional 
equilibrium) $x=1.2\times10^{-3}$, and ionization rate 
$\dot{N}_\gamma=5\times10^{48}$ photons\,s$^{-1}$. 
For these parameters the recombination time is 
$t_{\rm rec}=3.86\times10^{15}\,\rm s=122.4$ Myr. Assuming a recombination 
rate $\alpha_B(T)=2.59\times10^{-13}\rm\,cm^{3}s^{-1}$ at $T=10^4$ K, then
$r_S=5.4$ kpc. Note that the value of $r_S$ is actually independent from the use of the OTS
or Case B approximation (see Test 7).

\begin{figure}
\begin{center}
\includegraphics[totalheight=0.27\textheight]{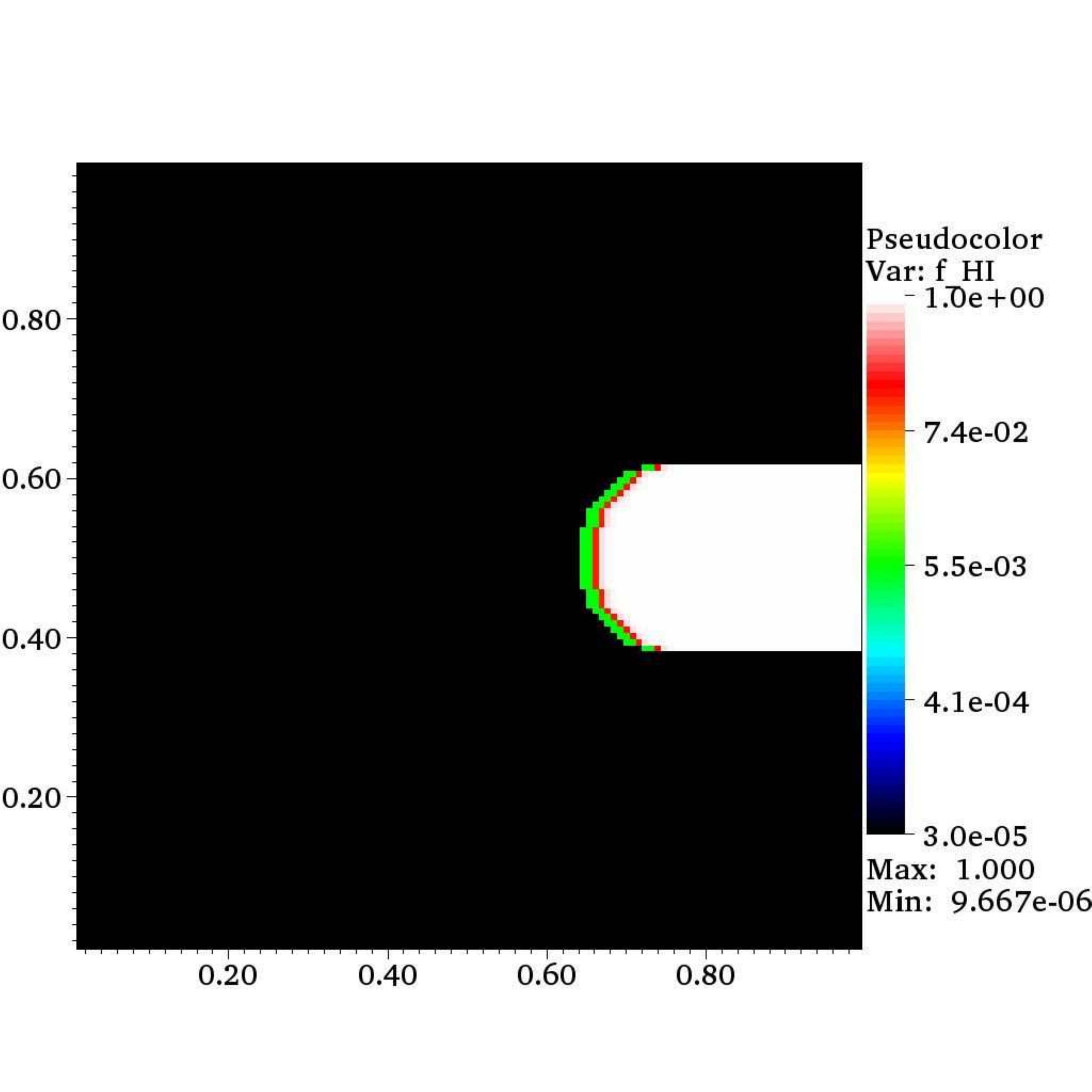}
\includegraphics[totalheight=0.27\textheight]{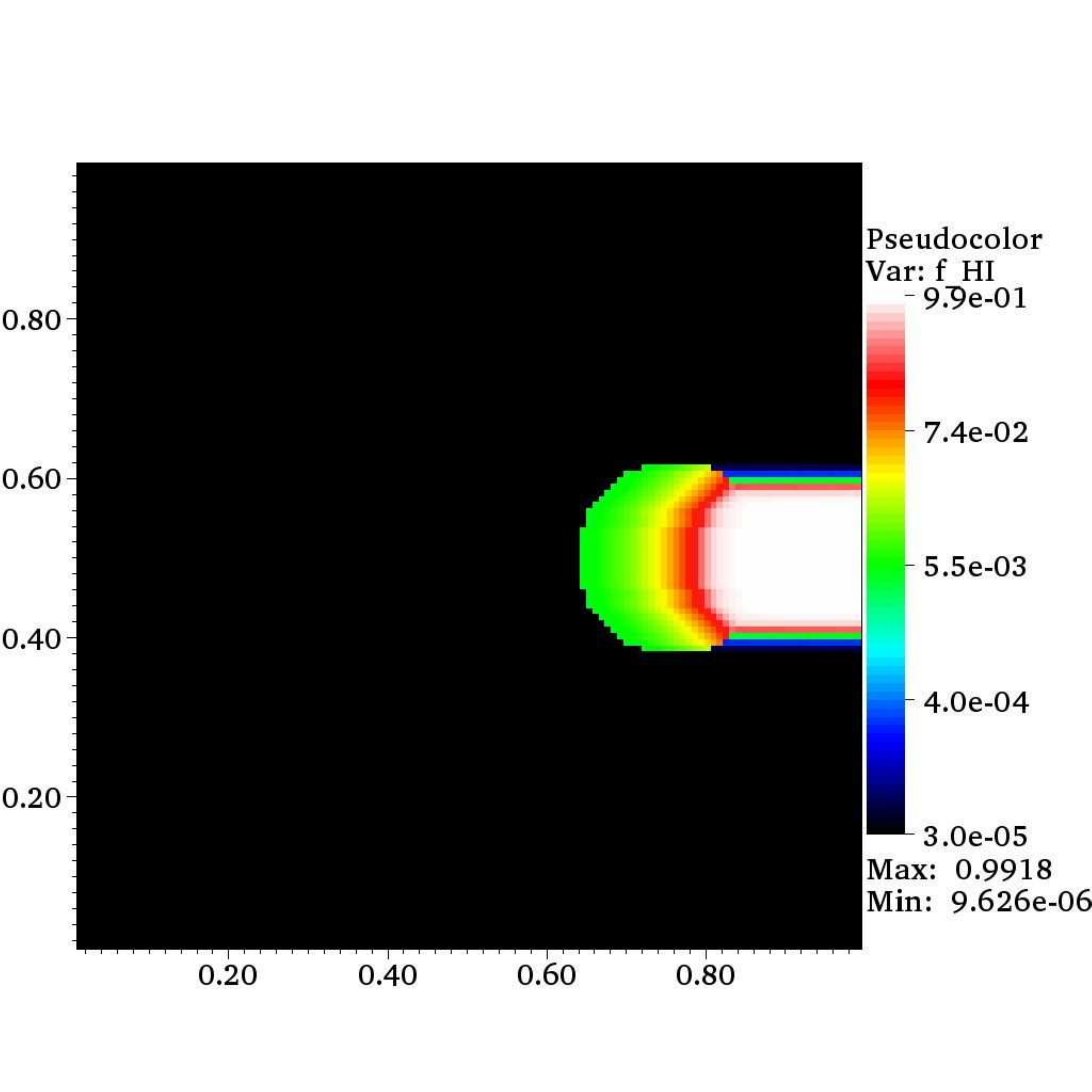}
\caption[Test 3: HI image slice]
{Test 3: image slices of the HI fraction at coordinate z=0.5 (box units), at times $t=1$ Myr (top panel)
and $t=15$ Myr (bottom panel).}
\label{T3HI_ima}
\end{center}
\end{figure}
\begin{figure}
\begin{center}
\includegraphics[totalheight=0.27\textheight]{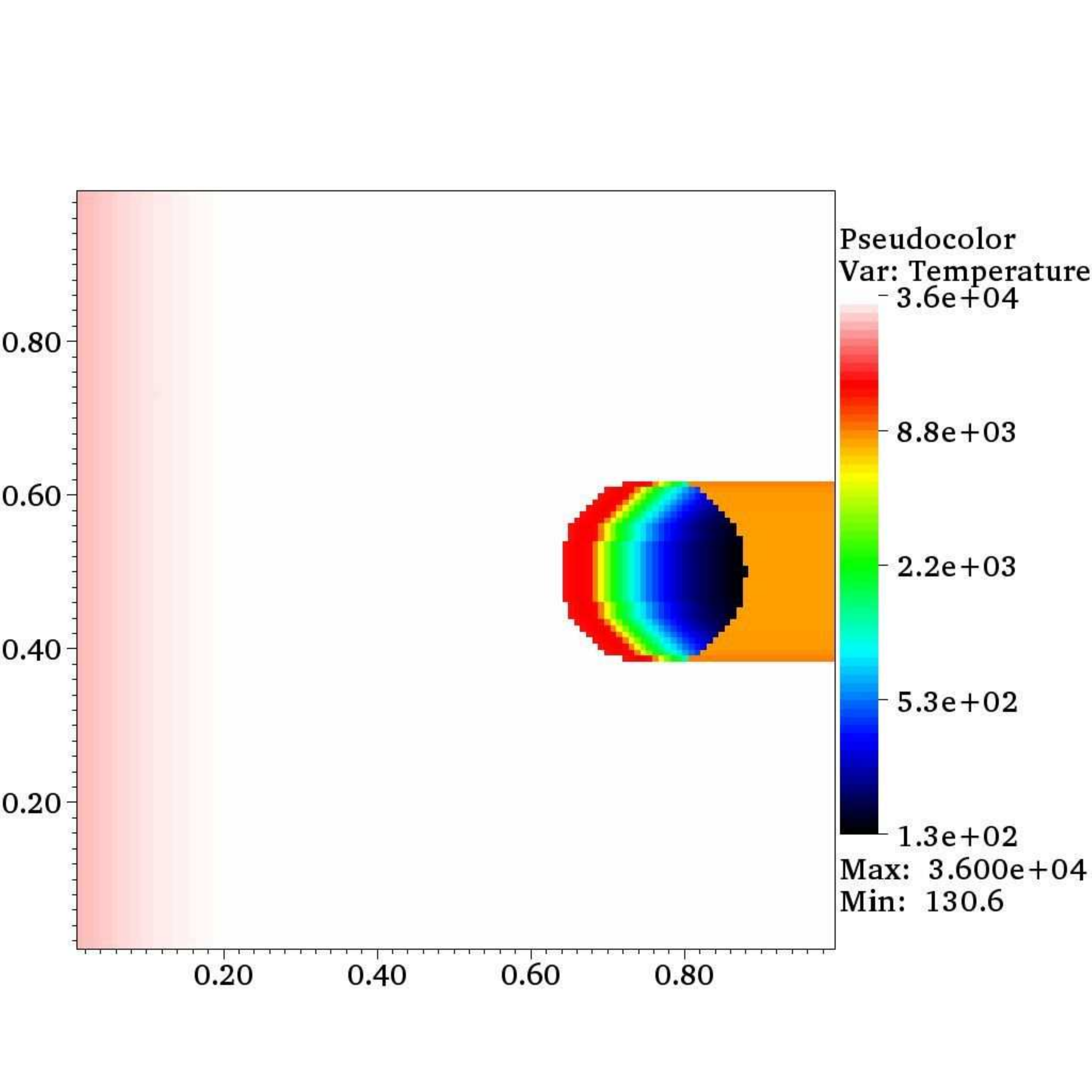}
\includegraphics[totalheight=0.27\textheight]{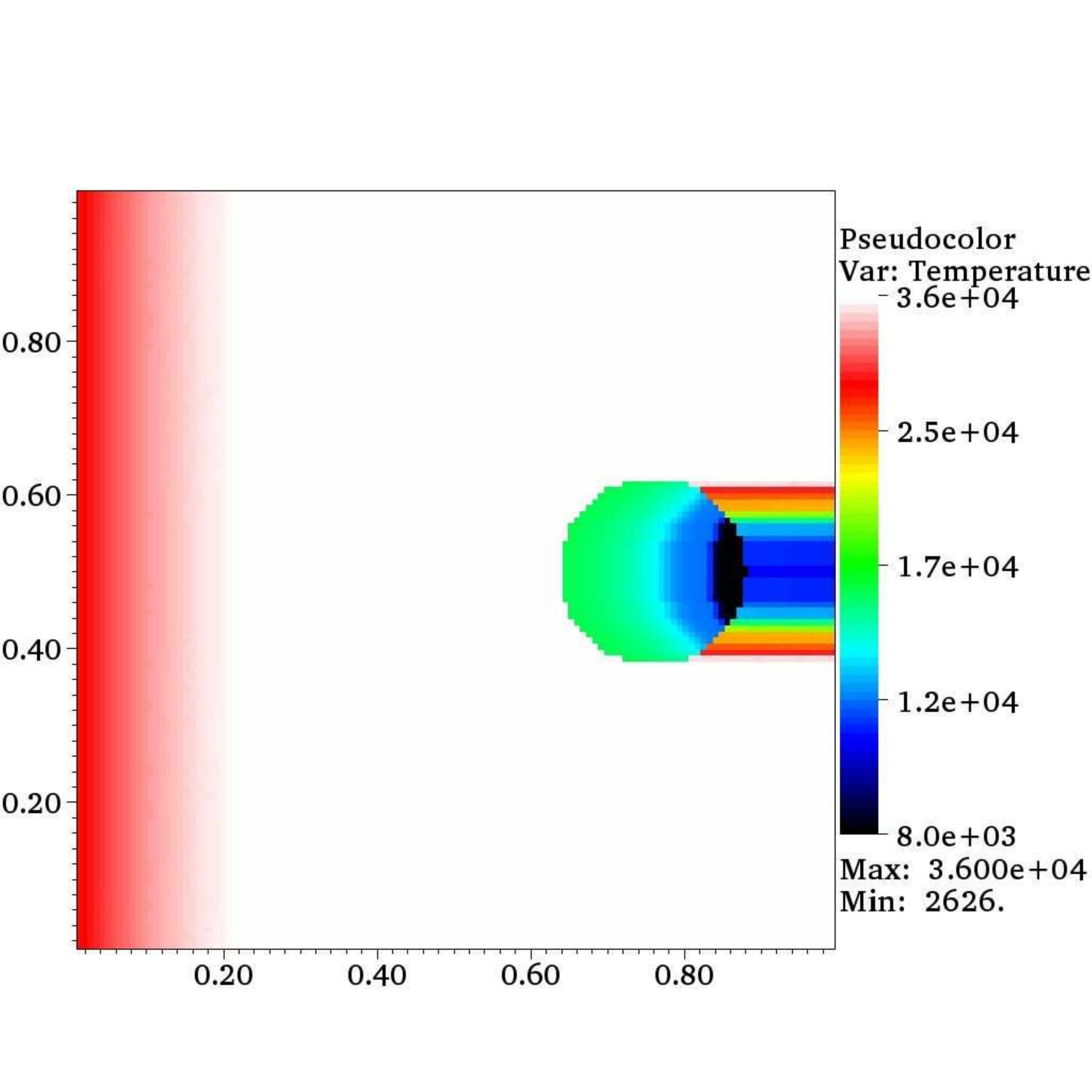}
\caption[Test 3: temperature image slice]
{Test 3: image slices of the gas temperature at coordinate z=0.5 (box units), at times $t=1$ 
Myr (top panel) and $t=15$ Myr (bottom panel).}
\label{T3Temp_ima}
\end{center}
\end{figure}

In Figure \ref{T1}, we show the images of the HI fraction in the plane y=0 and z=0 at time
$t=500$ Myr, when the equilibrium Str\"omgen sphere is reached. The HII region
is nicely spherical in both the planes, demonstrating that our algorithm
produces an uniform coverage of the solid angle around the source.
In Figure \ref{T1f8}, we plot the spherically averaged radial profiles of the ionized ($x$)
and neutral fraction (1-$x$) at times $t=30$ and 500 Myr.
The thickness of the transition between the HII and HI regions is in agreement
with both analytical and numerical expectations from most of the other codes in I06. 
In particular, thanks to our new adaptive algorithm (that ensures the convergence of the
radiation field in any cell), the I-front thickness does not suffer
from diffusive effects shown by other ``classical'' Monte Carlo methods
(cfr., e.g., the results of \texttt{CRASH} in I06).  
Finally, in Figure \ref{T1f7}, we show the time evolution of the I-front position
(defined as the point of 50\% ionization). The code tracks the I-front
correctly, with the position never varying by more than few percent from the
analytical solution.

\begin{figure}
\begin{center}
\includegraphics[totalheight=0.27\textheight]{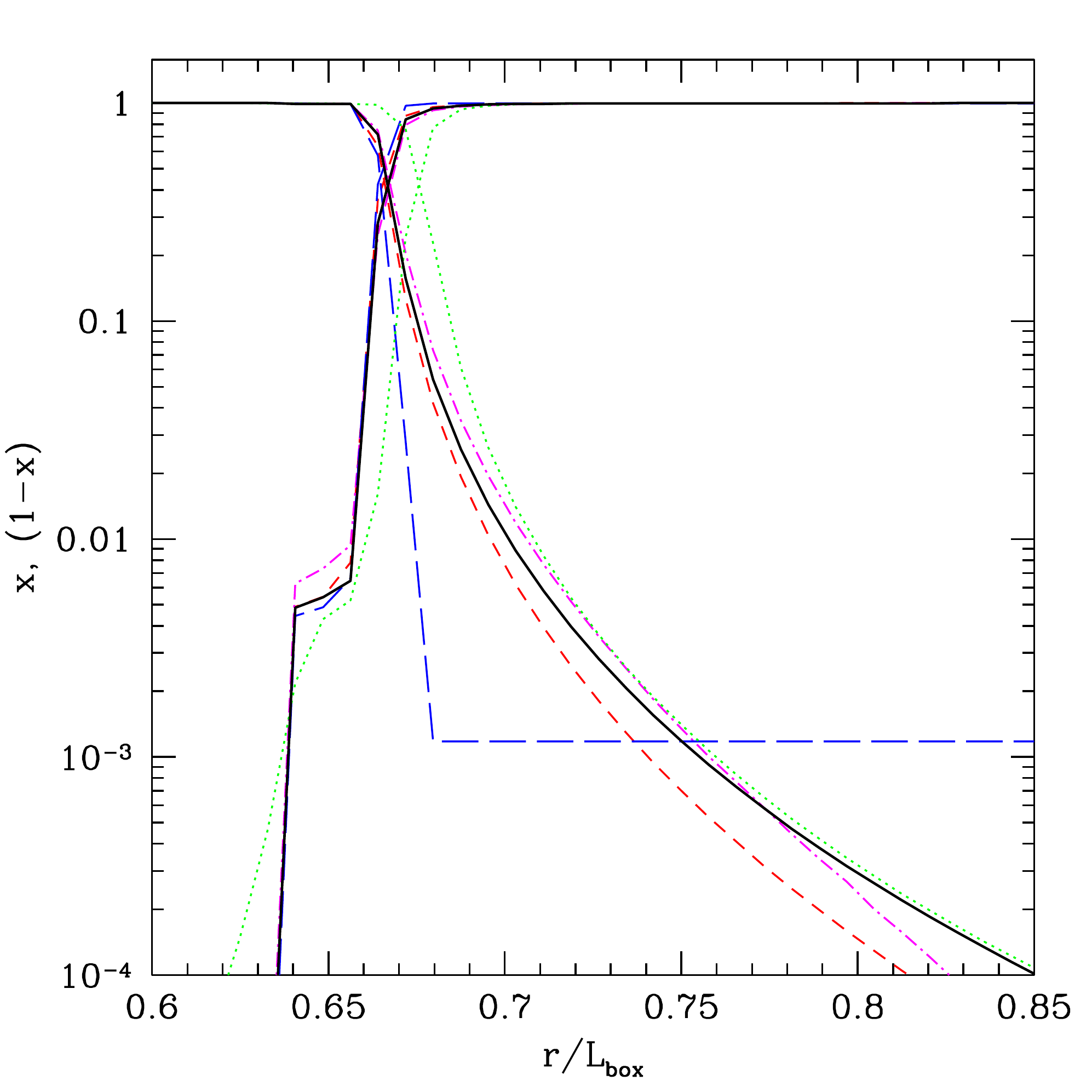}
\includegraphics[totalheight=0.27\textheight]{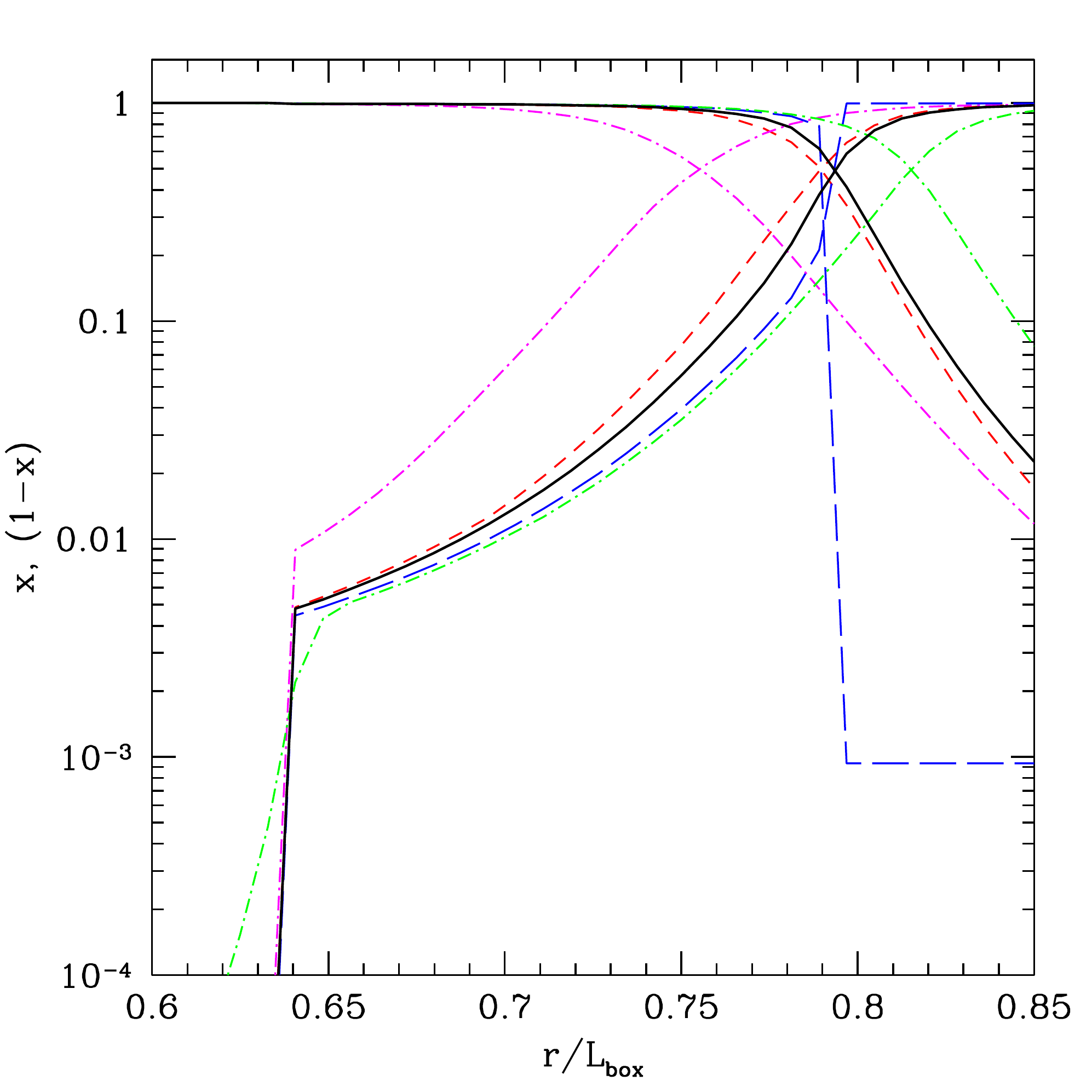}
\caption[Test 3: HI image slice]
{
Test 3: line cuts of the ionized and neutral fraction along the axis of symmetry through the centre of the clump at times $t=1$ Myr
(top panel) and $t=15$ Myr (bottom panel). The results from \cname are presented as solid black lines. The other
lines represent the results from a sample of four different codes 
taken from the RT code comparison project (I06),  see caption in Figure \ref{T2f16}.
}
\label{T3HI_prof}
\end{center}
\end{figure}
\begin{figure}
\begin{center}
\includegraphics[totalheight=0.27\textheight]{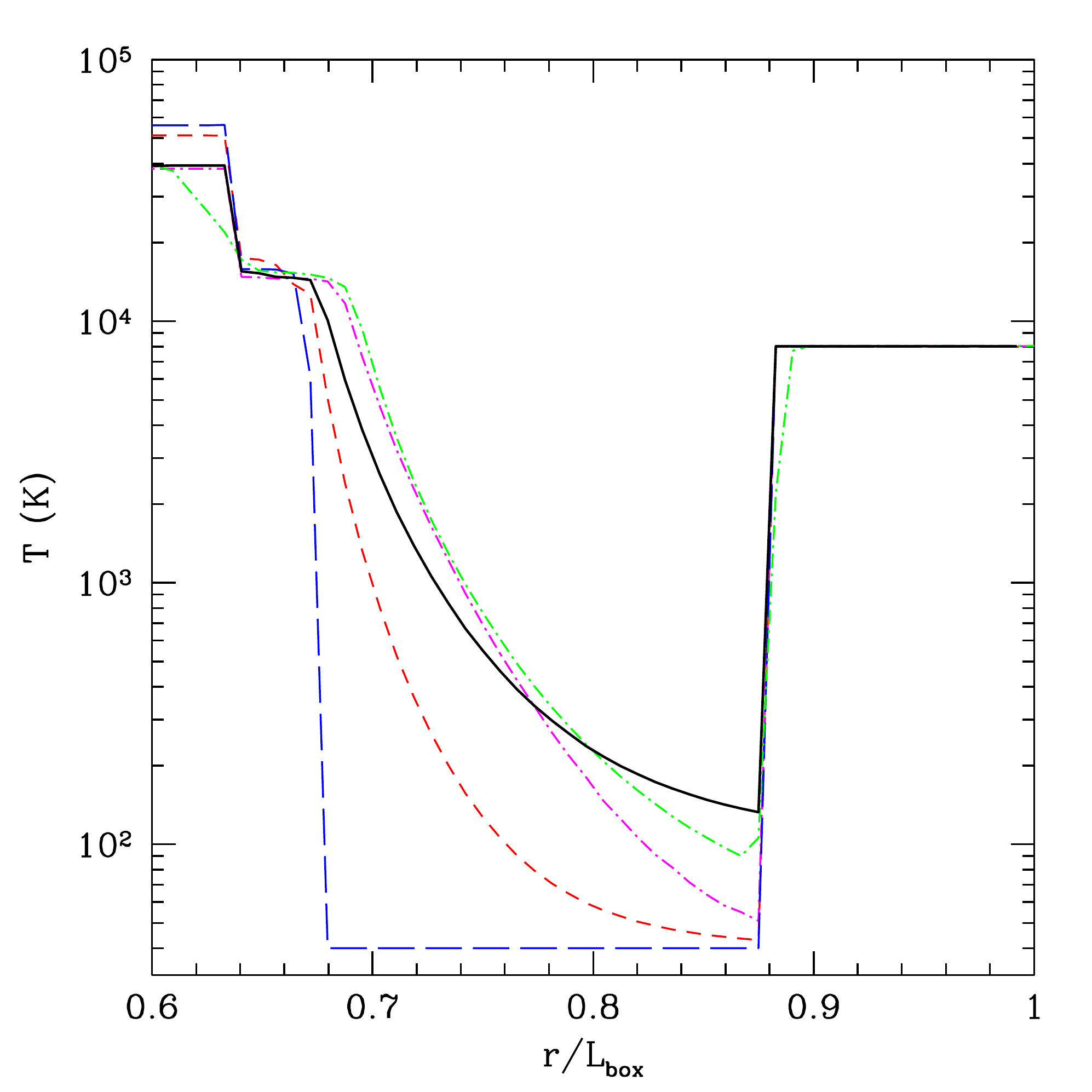}
\includegraphics[totalheight=0.27\textheight]{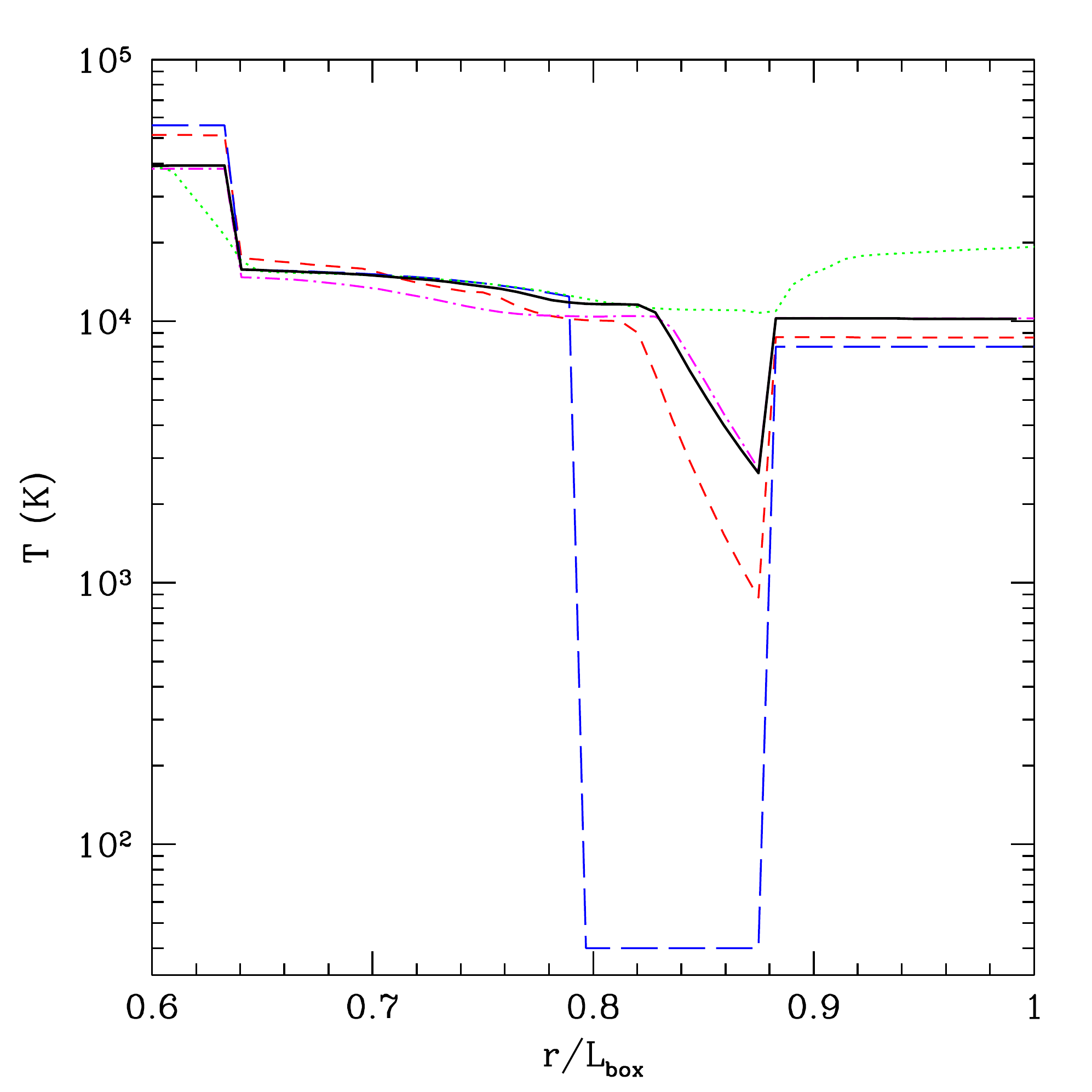}
\caption[Test 3: temperature image slice]
{
Test 3: line cuts of the temperature along the axis of symmetry through the centre of the clump at times $t=1$ Myr
(top panel) and $t=15$ Myr (bottom panel). The results from \cname are presented as solid black lines. The other
lines represent the results from a sample of four different codes 
taken from the RT code comparison project (I06), see caption in Figure \ref{T2f16}. 
}
\label{T3Temp_prof}
\end{center}
\end{figure}

\subsection{Test 2: HII region expansion with temperature evolution}

In this test we use the same parameters of Test 1, but now the ionizing source
has a $10^5\mathrm{K}$ black-body spectrum and we allow the gas temperature
to evolve. Initially, the gas is fully neutral with a temperature T=100 K.
There are no analytical solutions for this test, therefore we compare our
results to what obtained by the other codes in I06.

In Figure \ref{T2HI}, we show the images of the 
neutral hydrogen fraction (on the $z=0$ plane) at times
$t=10$ and $t=100$ Myr. The spherically averaged HI profiles, for the same
time-snapshots, are presented in Figure \ref{T2f16} (black, solid lines),
together with a sample of the results obtained by the other codes in I06 
(see caption in the Figure).
The overall size of the HII region and the internal structure
agree very well.
The temperature images and the spherically averaged profiles 
at times $t=10$ and $t=100$ Myr are presented in Figures \ref{T2Temp} and \ref{T2f17}. 
Also in this case, the resulting temperature structure agree well with most of the other codes in I06. In particular, we are able to obtain 
a large pre-heated region without any sign of spatial anisotropy, since our algorithm does not suffer from the under-sampling
the radiation field (like in a ``classical'' Monte Carlo), even a large distances from the sources.

\begin{figure}
\begin{center}
\includegraphics[totalheight=0.3\textheight]{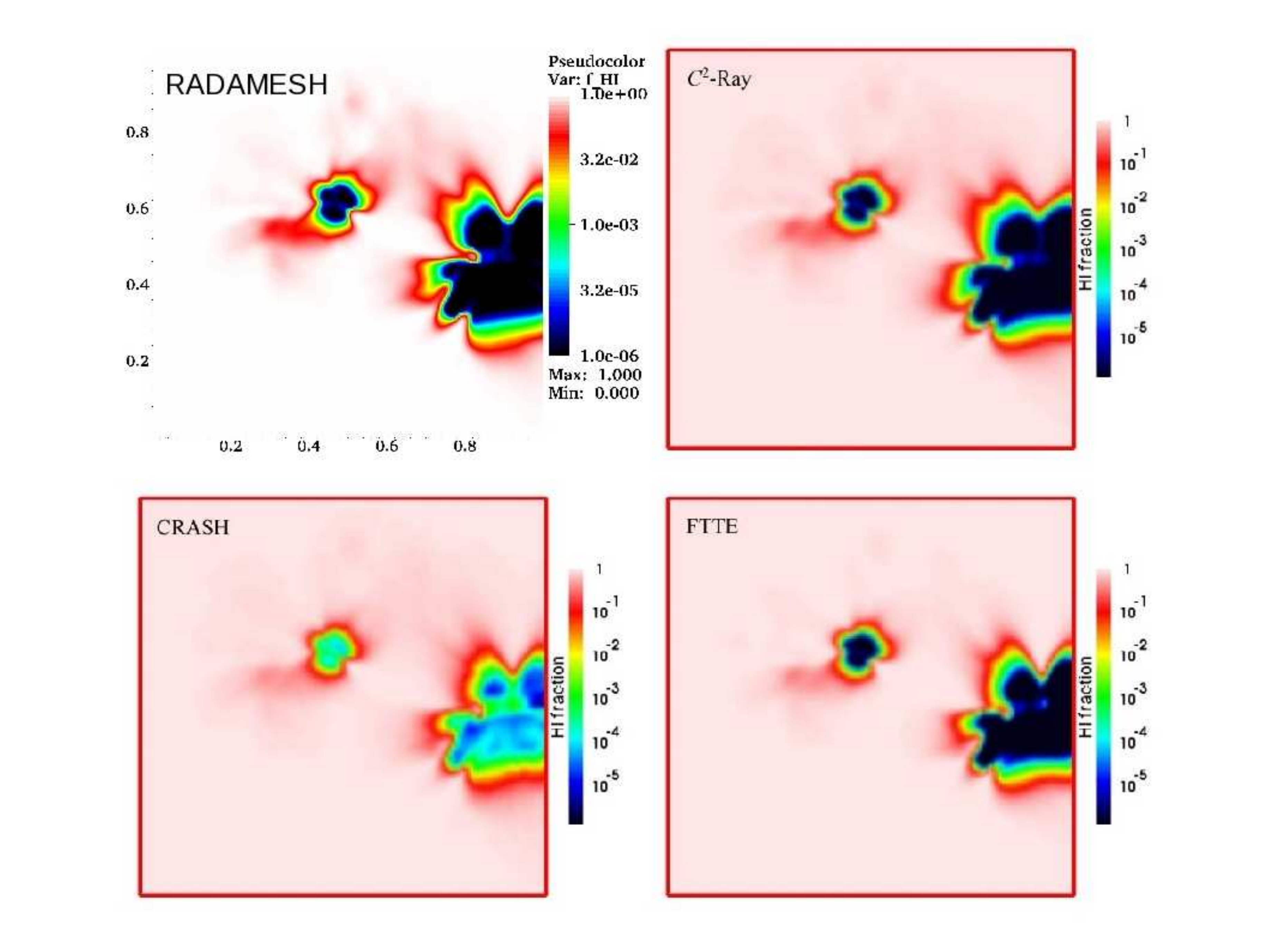}
\caption[Test 4: HI image slice at 0.05 Myr]
{
Test 4: image slices of the HI fraction at coordinate z=0.5 (box units) and time $t=0.05$ Myr obtained by \cname (top-left panel)
and by three of the four codes that performed the same tests in I06, in particular: \texttt{CRASH} (bottom-left panel), \texttt{C$^2$-RAY} (top-right panel),
and \texttt{FTTE} (bottom-right panel).
}
\label{T4HI_005}
\end{center}
\end{figure}
\begin{figure}
\begin{center}
\includegraphics[totalheight=0.3\textheight]{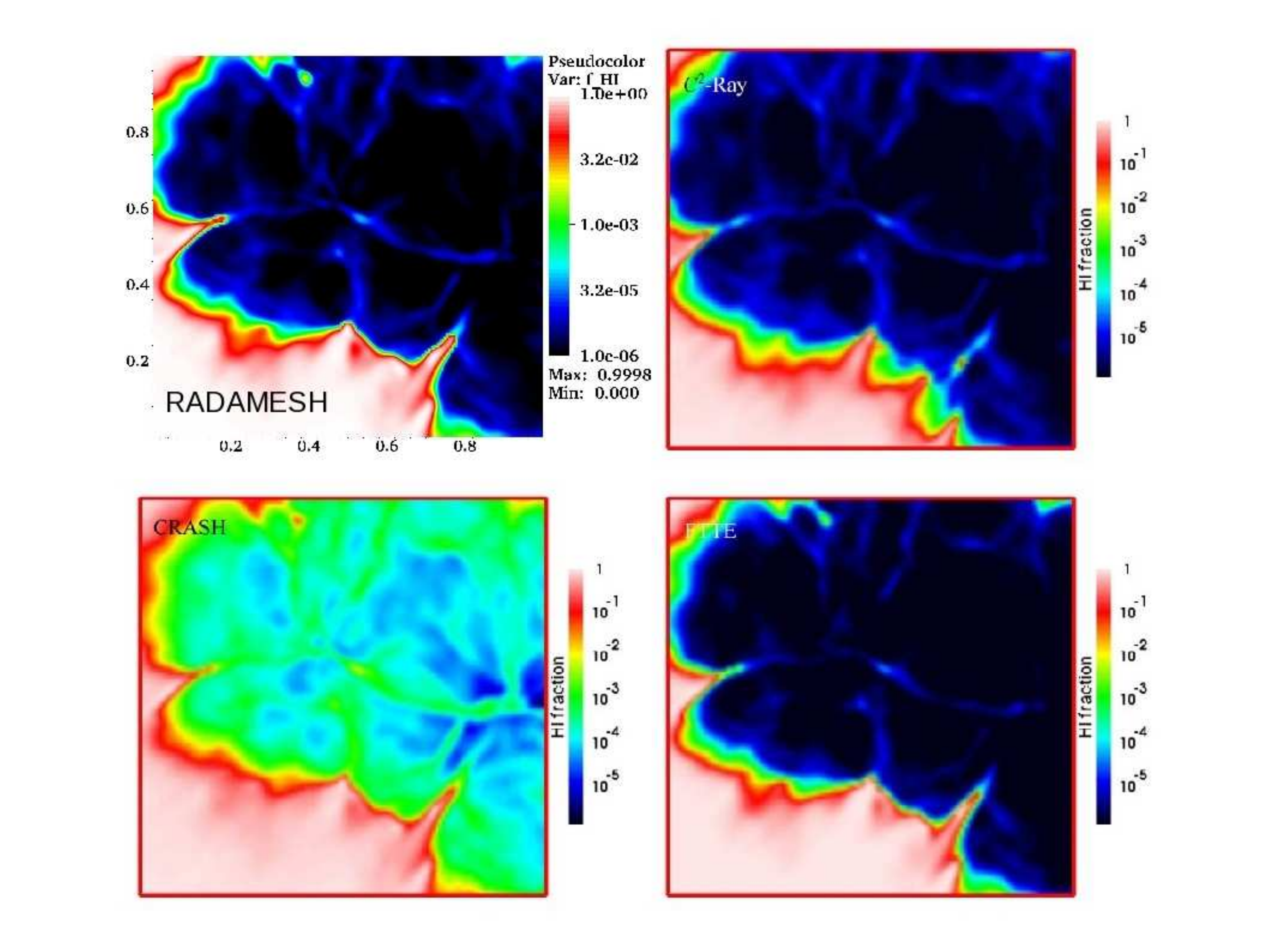}
\caption[Test 4: HI image slice at 0.2 Myr]
{
Test 4: image slices of the HI fraction at coordinate z=0.5 (box units) and time $t=0.2$ Myr obtained by \cname (top-left panel)
and by three of the four codes that performed the same tests in I06 (see caption in Figure \ref{T4HI_005}).
}
\label{T4HI_02}
\end{center}
\end{figure}

\begin{figure}
\begin{center}
\includegraphics[totalheight=0.3\textheight]{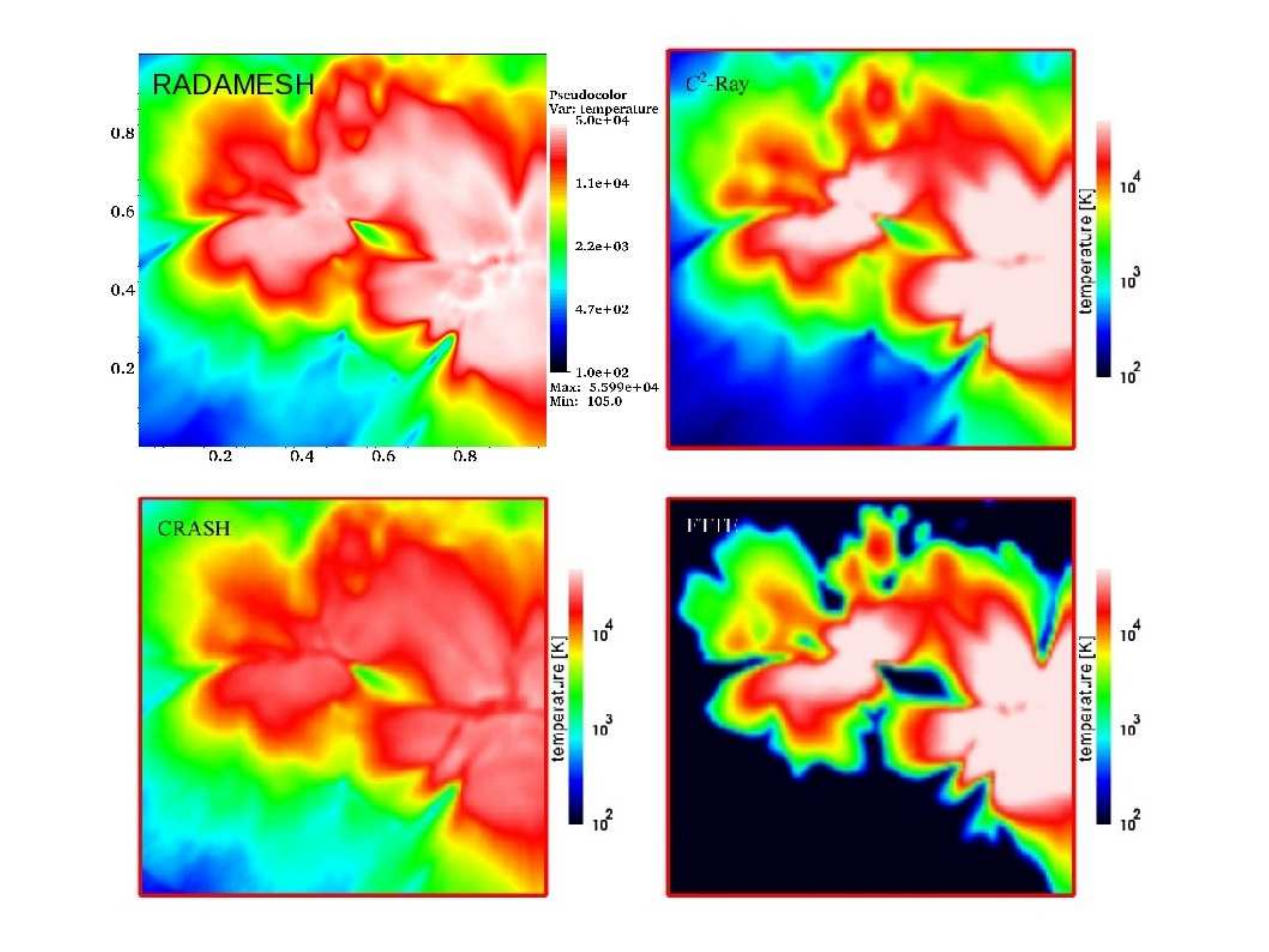}
\caption[Test 4: temperature image slice at 0.05 Myr]
{
Test 4: image slices of the gas temperature at coordinate z=0.5 (box units) and time $t=0.05$ Myr obtained by \cname (top-left panel)
and by three of the four codes that performed the same tests in I06 (see caption in Figure \ref{T4HI_005}).
}
\label{T4Temp_005}
\end{center}
\end{figure}

\begin{figure}
\begin{center}
\includegraphics[totalheight=0.3\textheight]{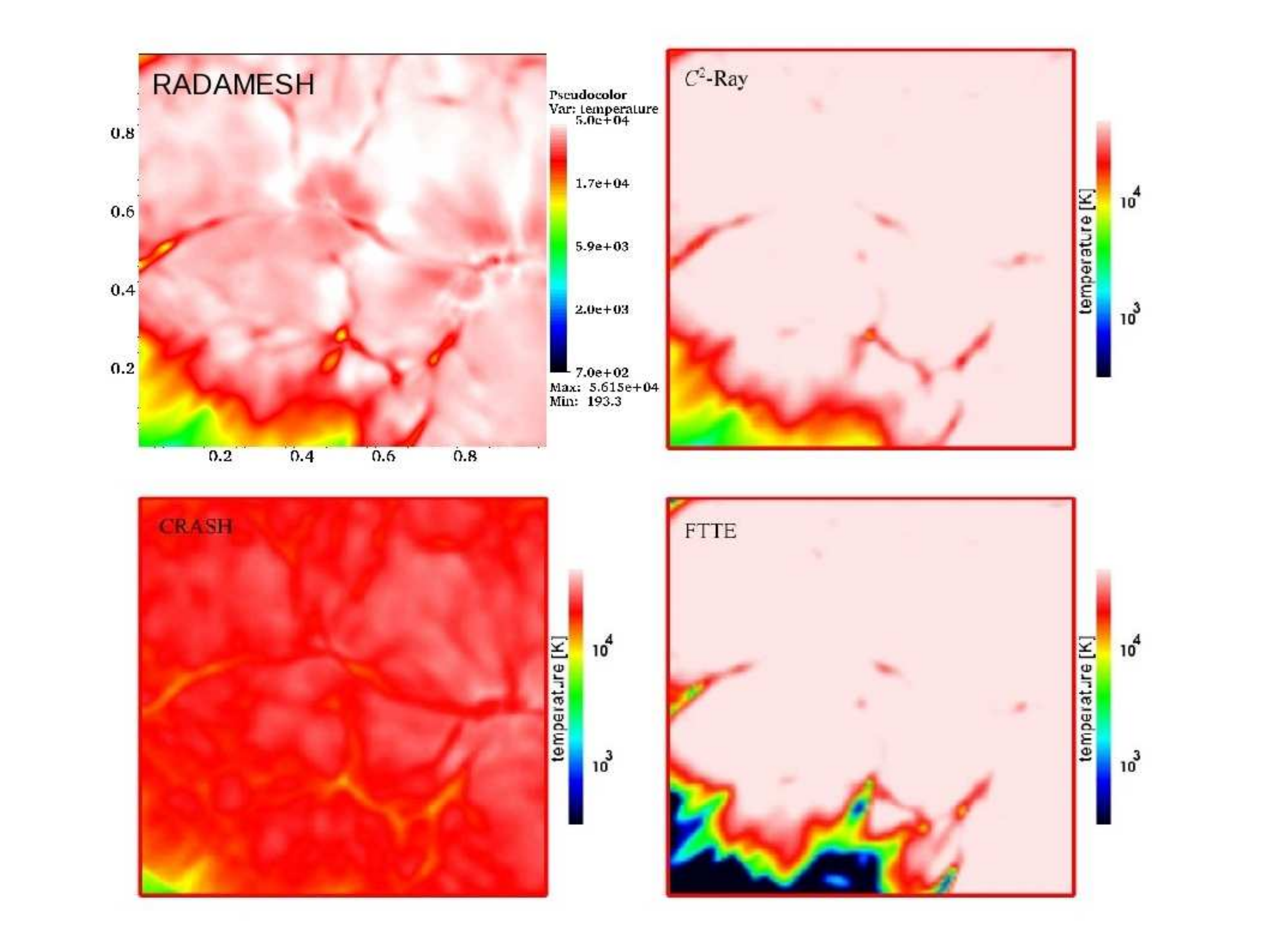}
\caption[Test 4: temperature image slice at 0.2 Myr]
{
Test 4: image slices of the gas temperature at coordinate z=0.5 (box units) and time $t=0.2$ Myr obtained by \cname (top-left panel)
and by three of the four codes that performed the same tests in I06 (see caption in Figure \ref{T4HI_005}).
}
\label{T4Temp_02}
\end{center}
\end{figure}

\subsection{Test 3: I-front trapping}

This test verifies that the propagation of an I-front within a dense clump is slowed down at the point
of being stopped (\emph{trapped}) and the production of a correct shadowing effect (in both the ionization
and the temperature state) behind the clump. 
We use the same set-up as described by I06: a spherical (hydrogen-only) uniform cloud of radius $r_{clump}=0.8$ kpc,
is located at the position $(5,3.3,3.3)$ kpc within a box of length $L=6.6$ kpc. The hydrogen
number density and the initial temperature outside of the clump are, respectively,  $n_{out}=2\times10^4$ cm$^{-3}$ 
and $T^{init}_{out}=8000$K, while inside the clump we have $n_{in}=200n_{out}=0.04$ cm$^{-3}$ and $T^{init}_{clump}=40$ K.
The radiation has a black-body spectrum with $T_{eff}=10^5$ K and a constant ionizing photon flux $F=10^6$ s$^{-1}$ cm$^{-2}$,
incident to the $y=0$ box side.

 Given these parameters, we expect that the I-front should be trapped slightly beyond the clump centre, as discussed
in I06. In Figure \ref{T3HI_ima}, we show slices of the gas neutral fraction at the box mid-plane $z=3.3$ kpc (passing through the
centre of the clump) at times $t=1$ Myr (top panel) and $t=15$ Myr (bottom panel). The corresponding temperature
slices are presented in Figure \ref{T3Temp_ima}.
At $t=1$ Myr, the I-front is not yet trapped and it is still moving supersonically
from the edge of the clump. The images show that the shadow is sharp and produced correctly behind the clump.
As expected, the I-front is trapped slightly beyond the clump centre at $t=15$ Myr.
The position of the I-front and the HI profiles inside the clump, presented in Figure \ref{T3HI_prof},
agree well with the other codes in I06, altough, especially at later times, the results are rather code dependent.
The same is true for the temperature profiles, as shown in Figure (\ref{T3Temp_prof}).

\subsection{Test 4: Multiple sources in a ``cosmological'' density field}

In this test, we follow the propagation of the Ionization-fronts from multiple sources
in a static cosmological density field. The initial conditions
for the density and source spatial distribution/luminosities are 
provided by I06. In particular, we use a time-slice at
$z=9$ from a hydro-simulation with a box size of $0.5\,h^{-1}$ comoving Mpc
and $128^3$ cells. The initial temperature is fixed to T=100 K everywhere.
The ionizing sources correspond to the 16 most massive halos in the box,
with a luminosity proportional to the halo mass and a black-body spectrum with
$T=10^5$ K (see I06 for more details).

In Figures \ref{T4HI_005} and \ref{T4HI_02}, we present slices of neutral hydrogen fraction cut through the simulation
box at coordinate $z=0.5$, in box units, at time $t=0.05$ Myr and $t=0.2$ Myr, respectively. 
Comparing our results with three of the four codes that performed this test in I06
(\texttt{C$^2$-Ray}, \texttt{CRASH} and \texttt{FTTE}), we find a general agreement, although all the codes
produce somewhat different morphologies. 
This general agreement is confirmed also examining the
temperature slices in Figures \ref{T4Temp_005} and \ref{T4Temp_02}. 
Here the differences between the codes presented in I06 is higher
and due to different assumption about spectral hardening and partially to the different
algorithm used. The spectral hardening effect, namely the increased temperature
in the regions that are still mostly neutral, is traced in detail by \cnamens.

\begin{figure}
\begin{center}
\includegraphics[totalheight=0.35\textheight]{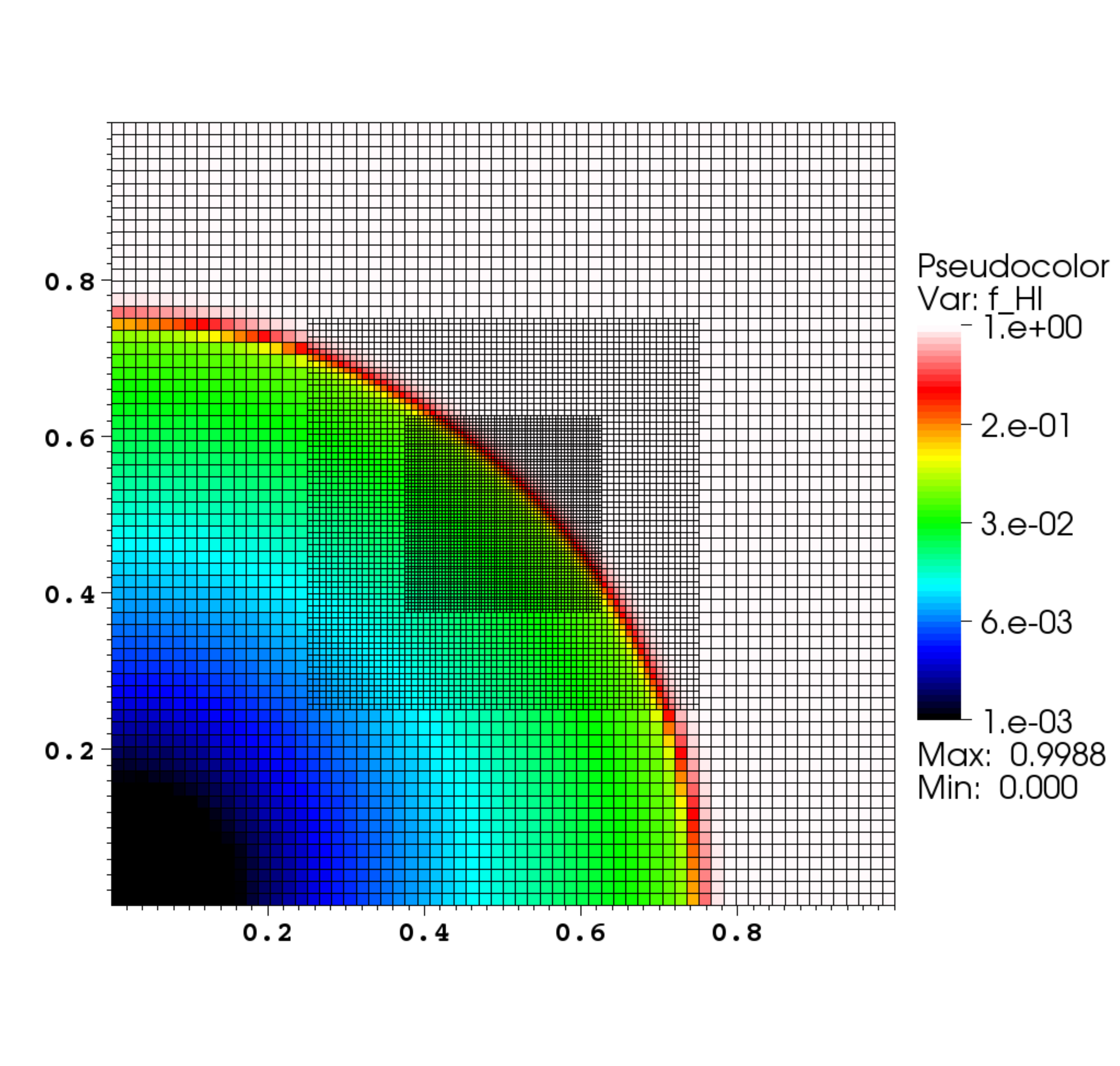}
\caption[Test 5: multi-grid structure]
{Test 5: (Ionization-front expansion in a static multi-mesh) HI fraction
image slice at time $t=200$ Myr and coordinate $z=0.5$ (box units).
The multi-mesh structure in the plane $z=0.5$ is overlaid.    }
\label{T5mg}
\end{center}
\end{figure}

\subsection{Test 5: static multi-mesh}

With this test, not present in the original set of I06,
we show and verify the multi-grid capability of \cname in the case of a static multi-mesh. 
In particular we use the
same initial condition of Test 1 changing the original grid with
three nested, concentric grids with $64^3$ cells, corresponding to two levels
of refinement (with a factor two and four increased resolution with respect to the base mesh).
 With this configuration, the centre of the box has a resolution equivalent
to a $256^3$ cells grid. We limit the number of levels and the resolution of the
base grid in this test for illustrative purposes.  
The source position is $(x=0,y=0,z=0.5)$ in box units.
In Figure (\ref{T5mg}), we show the image of HI fraction and the
computational meshes corresponding to the slice 
at coordinate $z=0.5$ and time $t=200$ Myr.
As we can see from the image, the front is tracked 
very well despite of the different grids (and resolution), with no spurious effect 
introduced by the multi-grid structure.

\subsection{Test 6: adaptively refining multi-mesh}

In the original set of tests in I06, the initial conditions (e.g., box size, hydrogen density, source luminosity)
have been chosen in such a way to properly resolve the I-front with a single, uniform grid
of $128^3$ cells. For example, in the final test of I06, Test 4  (multiple sources in a ``cosmological'' density field),
the box size is fixed to $0.5\,h^{-1}$ comoving Mpc, corresponding to $\sim70$ physical kpc at the simulation redshift
($z=9$) for $h=0.7$. 
However, in a more realistic cosmological situation, e.g. for the study of the reionization process or the expansion of the
I-front around a bright, high-redshift QSO, the box size must be at least 2 orders of magnitude larger than Test 4 (see, e.g., Meiksin 2009).

\begin{figure*}
\begin{center}
\includegraphics[totalheight=0.55\textheight]{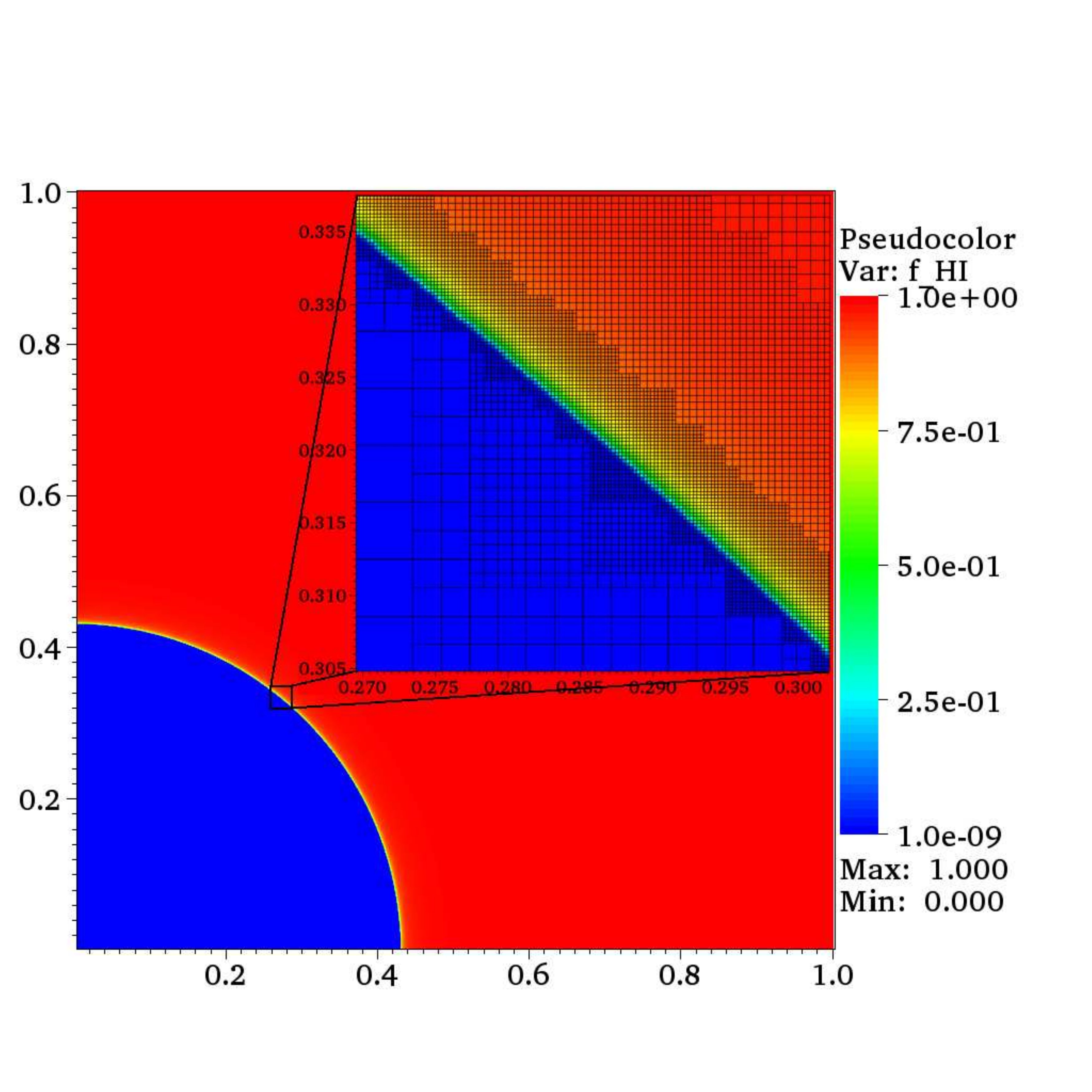}
\caption[Test 6: evolving multi-mesh structure]
{Test 6: (Ionization-front expansion in a evolving multi-mesh) HI fraction
image slice at time $t=100$ Myr and coordinate $z=0$ (box units).
The inset shows the evolved multi-mesh (composed of 5 levels of refinement) that closely follows the Ionization-front.   }
\label{T6image}
\end{center}
\end{figure*}

This test, not present in the original set of I06, demonstrate the ability of \cname to resolve the I-front
of a bright quasar in a large cosmological box, with the help of an adaptively evolving multi-mesh. In particular we show
that properly resolving the I-front is essential to accurately predict the temperature state of the gas,
 both in the I-front and in the QSO HII region. For this purpose we run a series of simulation increasing the maximum
level of refinement until convergence is reached for the temperature state of the gas. 
In the same spirit of Test 2, we use an uniform (hydrogen-only) medium with evolving temperature and we locate the source
at one corner of the box. The box size is $100$ comoving Mpc, the medium density is equal to the mean hydrogen density
of the universe and it evolves with redshift. The initial redshift is $z=7.5$, the total evolution time is $10^8$ yr 
(corresponding to a final redshift $z\sim6.7$). The source
has a power-law spectrum with a (frequency) spectral slope of $\alpha_{\nu}=1.7$, sampled in 60 logarithmically spaced bins
from 1 to 60 Rydberg. The total ionization rate is $N=10^{57}$ ph s$^{-1}$. These parameters are comparable to the
respective quantities associated with the observed high-redshift quasars in SDSS (e.g., Fan et al. 2006).  

For the assumed spectrum (which has a mean ionizing photon energy of $\sim2.4$ Ryd) the photon mean-free path 
is $\lambda^{\mathrm{HI}}_{\mathrm{mfp}}\sim6$ physical kpc at the initial redshift for
a fully neutral patch of gas composed of hydrogen only. 
 To resolve the I-front properly we would ideally need that $\lambda^{\mathrm{HI}}_{\mathrm{mfp}}$ is sampled by at least
two cells, i.e. a spatial resolution of $\sim3$ physical kpc. With an uniform grid this would correspond to a $4096^3$ cells mesh, 
effectively out of the reach for current computational facilities. Instead, we use a $128^3$ base grid and 5 additional levels of adaptive grid refinement
that follow the I-front expansion, effectively achieving the required spatial resolution. 

In Figure \ref{T6image}, we show an image slice of the hydrogen neutral fraction at $t=10^8$ yr (in the quasar rest-frame). 
The I-front scale is so small compared to the box size that
it visually appears perfectly thin in the large image. Zooming in by a large factor in a small region containing the I-front (see inset in the
same figure), and now the I-front thickness appears. In box units, the I-front size (as measured from the points where the neutral fractions are
0.1 and 0.9) is roughly $\sim0.0025$, i.e. $\sim32$ physical kpc, corresponding to $\sim5\lambda^{\mathrm{HI}}_{\mathrm{mfp}}$. 
Overlaid, we show the adaptively refined multi-mesh structure that closely follows the I-front, with the highest level of refinement ($l=5$) 
that encompasses the I-front itself. At this level of refinement, the I-front is fully resolved by at least 10 simulation cells.

\begin{figure}
\begin{center}
\includegraphics[totalheight=0.35\textheight]{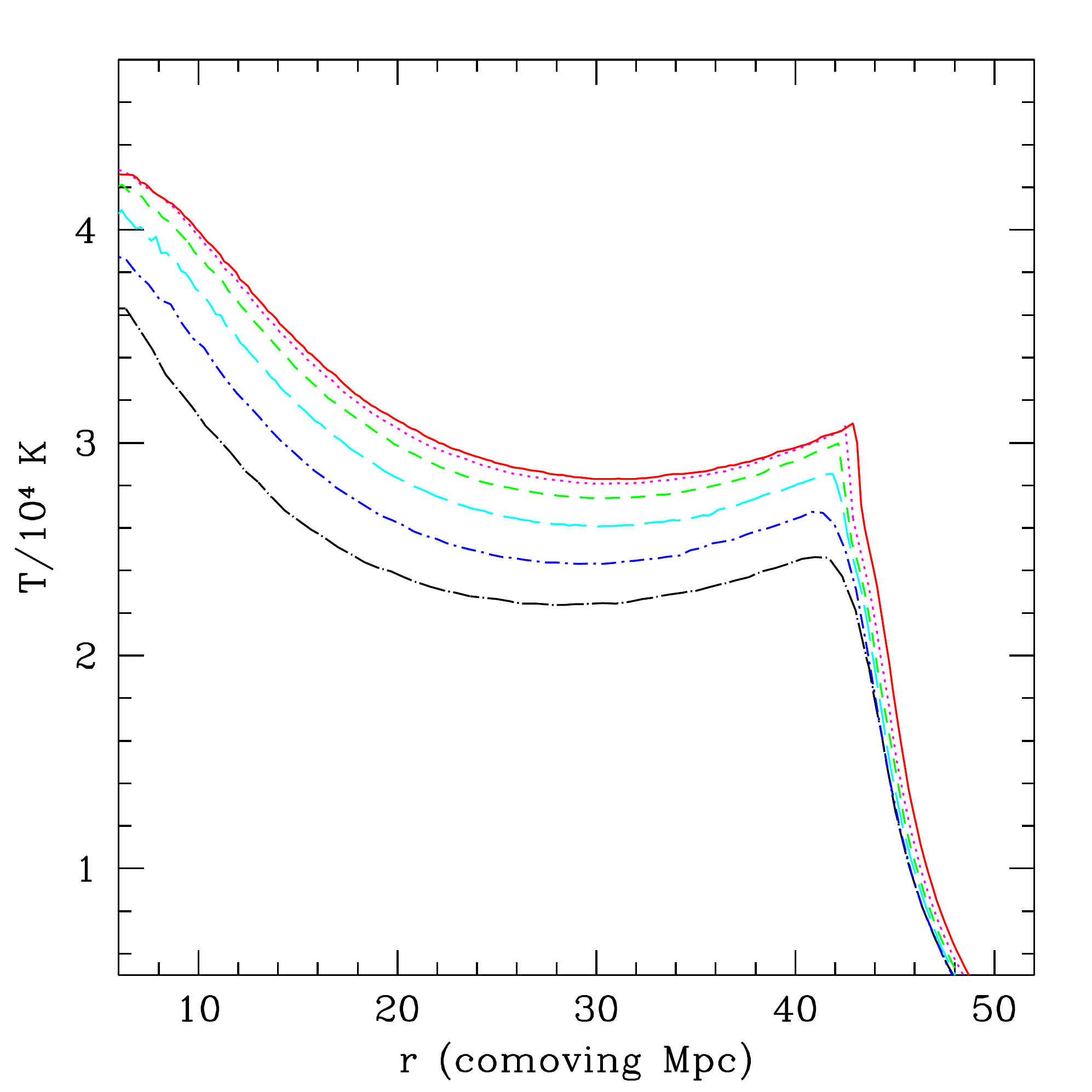}
\caption[Test 6: evolving multi-mesh structure]
{Test 6: (Ionization-front expansion in a evolving multi-mesh) Temperature profiles around a bright source (a QSO) in a 
large, cosmological box ($100$ comoving Mpc) and I-front adaptive mesh refinement. The QSO was turned on
at $z=7.5$ and keep it at a constant ionizing rate of $N=10^{57}$ ph s$^{-1}$ for $t=10^8$ yr (the final redshift is thus $z\sim6.7$). 
From bottom to top, the lines correspond to increasing maximum level
of refinement (from $l_{max}=0$ to $l_{max}=5$). The base grid is composed by $128^3$ cells. This plot clearly shows that the 
high-resolution achieved through AMR is essential to correctly resolve the I-front and, thus, to properly recover the gas temperature
profile around the QSO in a cosmological context.    }
\label{T6plot}
\end{center}
\end{figure}

 In Figure \ref{T6plot}, we show how resolving the I-front changes the predicted temperature profile inside the QSO HII bubble and in the I-front region.
From bottom to top, the lines represent the temperature profiles obtained varying the maximum level of refinement from $l_{max}=0$
(i.e., single uniform grid) to $l_{max}=5$. As expected, temperature convergence is reached for $l_{max}=4$, i.e. at spatial resolution
comparable or smaller than $\lambda^{\mathrm{HI}}_{\mathrm{mfp}}$, indicating that we are effectively resolving the I-front.
In fact, fully resolving the I-front improves the sampling of the spectral hardening 
(within the I-front itself) and a harder spectrum is able to produce a large \emph{increase} in the gas temperature,
as observed in Figure \ref{T6plot}\begin{footnote}{
Note that the current temperature of a cell within the HII region at $t=100$ Myr is mainly determined 
at an earlier epoch, i.e., when the cell was located within the I-front. Therefore, also the temperature increase
\emph{within} the HII region can be ascribed to the (I-front) spectral hardening effect. Numerical effects on the gas temperature 
due to the different resolution are negligible, as we show in Appendix B. 
}\end{footnote}.
The prediction of the correct temperature state in the I-front and in the QSO bubble is fundamental,
e.g., for the study of the reionization epoch with the near-zone Ly$\alpha$ forest (see e.g., Bolton et al. 2010) 
or with the I-front Ly$\alpha$ emission (Cantalupo et al. 2008). 

\subsection{Test 7: Diffuse ionizing emission from HII recombinations}

 We repeat Test 1 dropping the OTS approximation used in I06 
and including now the full radiative transfer of diffuse photons produced by hydrogen recombinations. 
Since all the cell in the Str\"omgren sphere may be now source of ionizing photons, we must change
the original test configuration, placing the (discrete) source at the centre of the box and increasing the box
size by a factor of two. Apart from these modifications, we use the same parameter set as in Test 1. 
 
 For a monochromatic spectrum, homogeneous medium and for a fixed temperature, we know that the radius of the Str\"omgren sphere
($R_s$)
obtained with the OTS (Case B) approximation must be equal to the one obtained with the full radiative
transfer of the diffuse radiation. This is a fundamental constraint deriving from photon conservation. 
Moreover, once the ionization equilibrium has been reached, 
there is an analytical approximation for the ratio between the diffuse and directional radiation field needed
to compensate, at a given radius $r$, the recombinations, as found by Ritzerveld (2005):

\begin{equation}\label{Ritz}
\frac{I^{\mrm{diff}}(r)}{I^{\mrm{dir}}(r)}\equiv \frac{\Gamma^{\mrm{diff}}(r)}{\Gamma^{\mrm{dir}}(r)}
=\left[1-\left(\frac{r}{R_s}\right)\right]^{1-\alpha/\alpha_{\mrm{B}}-1} \ ,
\end{equation} 

where the first equality is only valid for a monochromatic spectrum.
Note that eq. (\ref{Ritz}) is exact for an ``outward-only'' system, i.e. when the contribution to
$I^{\mrm{diff}}(r)$ is given by the recombinations at $r'<r$ (indeed, $I^{\mrm{diff}}(0)=0$ in Ritzerveld's solution). 
Therefore, we expect that this approximation should slightly underestimate $I^{\mrm{diff}}$ for $r\ll R_s$ (note, however that within 
this region the total radiation field will be dominated by $I^{\mrm{dir}}$). 

\begin{figure}
\begin{center}
\includegraphics[totalheight=0.27\textheight]{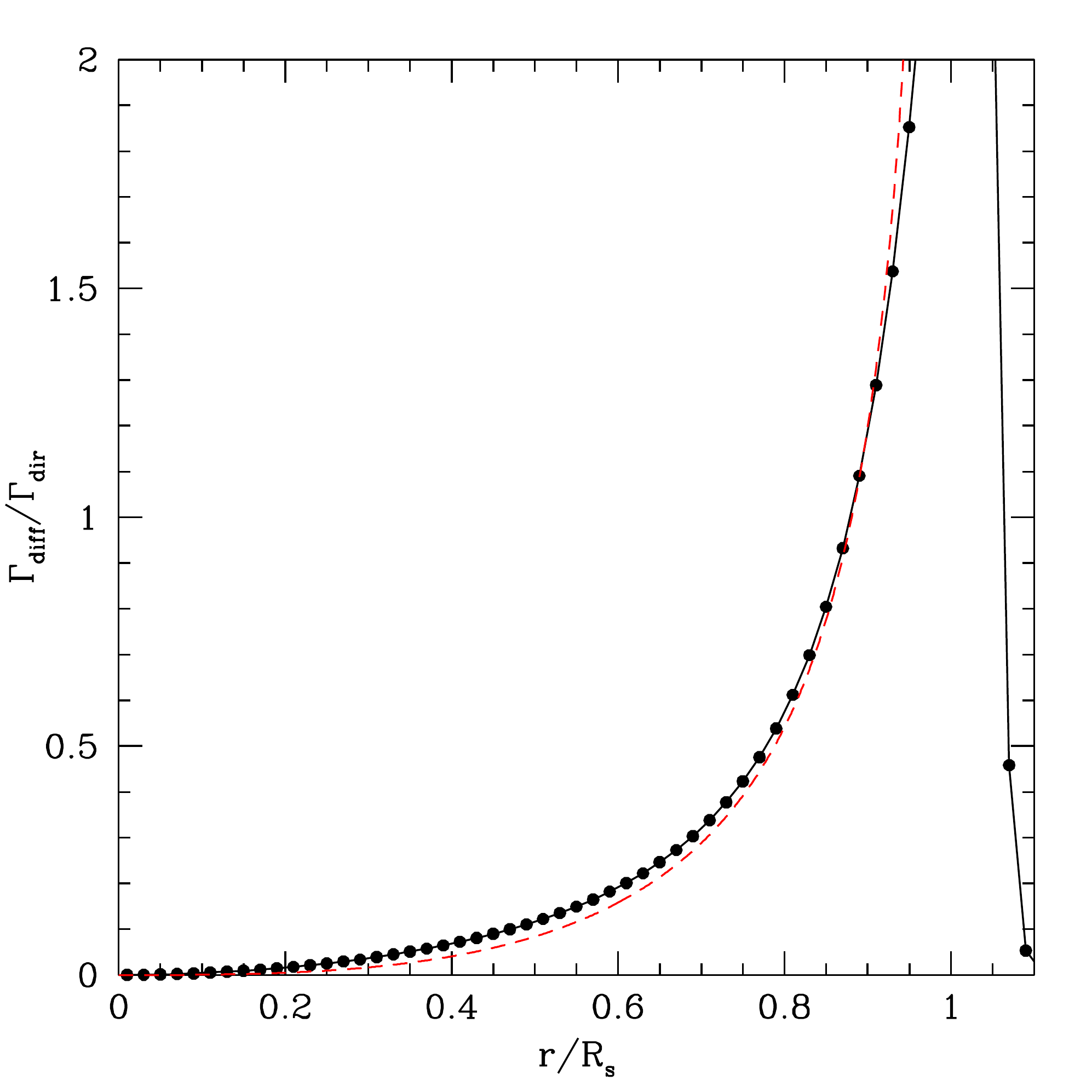}
\caption[Test 7: diffusion emission from HII]
{Test 7: (diffuse emission from HII) ratio between the ionization rate from diffuse HII recombinations 
and direct emission (monochromatic spectrum) 
as a function of distance from the
central source. For comparison, the analytical ``outward-only'' solution found by Ritzerveld (2005) is shown as a red dashed line.
Note that the analytical solution should became exact in proximity of $R_s$ (see text for details). 
In this test, the medium temperature has been fixed to $T=10^4$K    }
\label{GammaDiff}
\end{center}
\end{figure}

 In Figure \ref{GammaDiff}, we present the result obtained by \cname including the full radiative transfer of diffuse photons (black solid line).
In very good agreement with the analytical expectations (red dashed line), we found that the diffuse radiation field strength equals the directional
field from the central source at $r=0.87R_s$, confirming that diffuse radiation is dominant in the outer parts of the Str\"omgren sphere.
The full RT results produce, as expected, a slightly higher value of diffuse radiation in the central parts of the HII regions with
respect to the ``outward-only'' solution. Because of photon conservation, this is correctly balanced by a lower diffuse field at the
edge of the Str\"omgren sphere when compared to the analytical approximation.

\begin{figure}
\begin{center}
\includegraphics[totalheight=0.27\textheight]{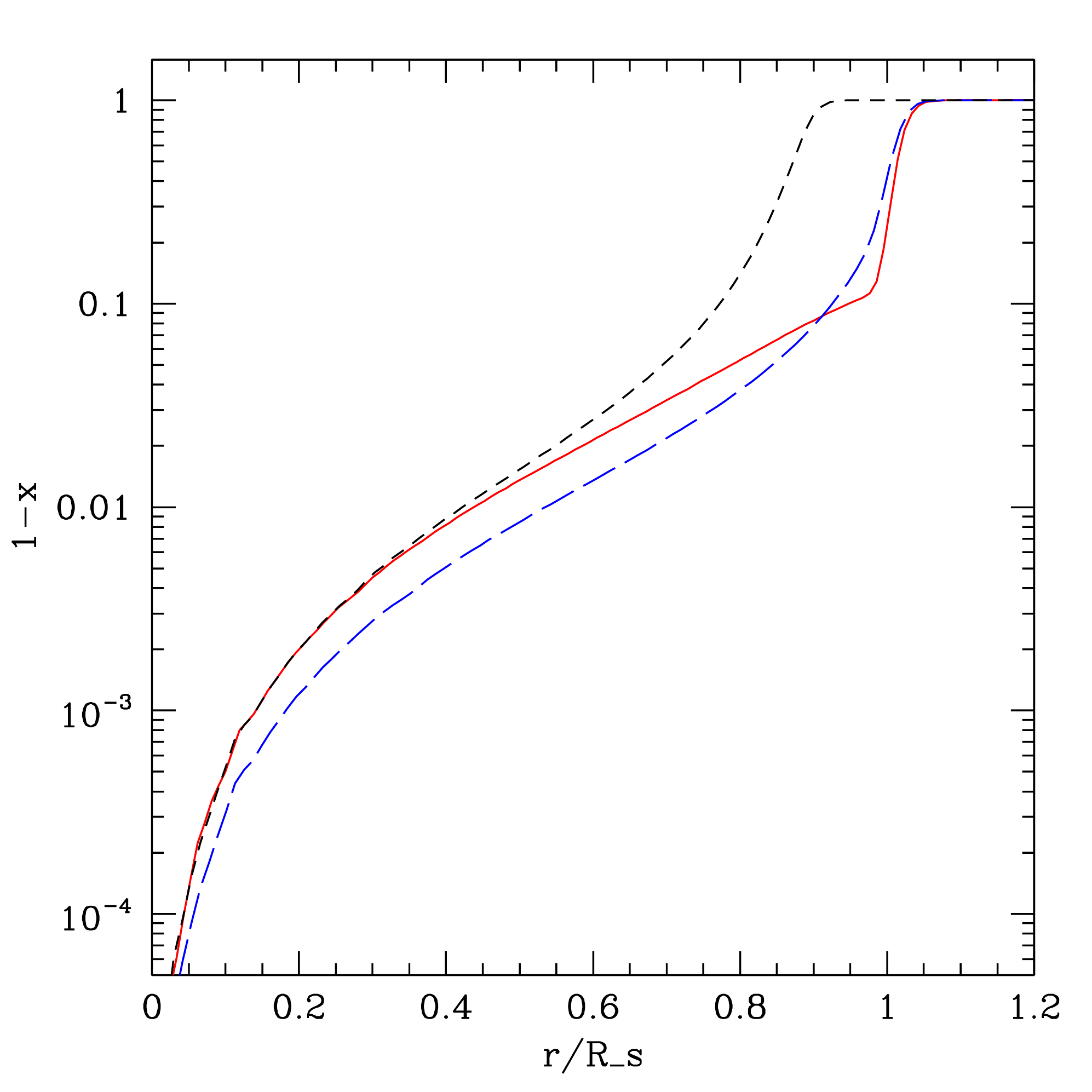}
\caption[Test 7: diffusion emission]
{Test 7: (diffuse emission from HII) HI fraction profile at $t=500$Myr for Case A (short dashed black line), ``on-the-spot'' Case B 
(long dashed blue line) 
approximations and including the full radiative transfer of diffuse emission from HII recombinations (solid red line). In this test, 
the medium temperature
has been fixed to $T=10^4$K.      }
\label{HIfracDiff}
\end{center}
\end{figure}

 In Figure \ref{HIfracDiff}, we show how the hydrogen neutral fraction profile ($1-x$) at ``equilibrium'' ($t=500$ Myr) 
is modified by the full radiative transfer
of diffuse photons (red solid line) with respect to the OTS (Case B) approximation (blue, long-dashed line) and the other extreme represented by 
the Case A approximation (black, short-dashed line). As expected, in the interior of the HII regions, where the gas is highly ionized,
the mean-free-path of the diffusion photons is very large and thus the profile obtained by the full RT closely follows the
one obtained assuming Case A. However, as the local opacity increases moving outward, diffuse photons start to be absorbed and the HI fraction
tends to become closer to the OTS case. In the outer parts of the HII region ($r>0.87R_s$), diffusion radiation becomes
the dominant component and the gas is more ionized than in the OTS Case B approximation. However, in accordance with photon conservation, the 
final size of the Str\"omgren sphere is very close to the one obtained with Case B: the recombination radiation escaped from the inner region
has been absorbed (creating an ``excess'') closer to the Str\"omgren radius, but the final photon balance is the same.

 As a remark, it is interesting to note that the OTS approximation is actually never valid inside the Str\"omgren sphere for this
very simple, standard test (apart, obviously, at the Str\"omgren radius). Actually, Case A is a much better approximation
if one is interested in the inner part of the HII region, or, analogously, in the early phases of the Ionization-front expansion.

\subsection{Test 8: Diffuse ionizing emission from HII, HeII and HeIII recombinations}\label{RecRadSec2}

 The previous test has been limited to the very simple case of a monochromatic spectrum (for both diffuse and direct radiation) and
a hydrogen only, fixed temperature medium. In this way, we were able to compare our results to the analytical prediction from photon conservation arguments.
However, the real effect induced by the full RT of recombination radiation is much more complex since its characteristic spectral energy distribution (SED)
is in general different from the SED of the discrete sources. Moreover, the presence of helium may have a profound impact also on the
neutral hydrogen distribution, given that both the bound-free and the Balmer continuum of HeI and HeII (together with two-photon and line emissions) 
are able to ionize HI.

 In this test, we verify these effects using the same configuration of Test 2, including now the full RT of recombination radiation
from hydrogen and helium. The only parameters, apart the inclusion of the diffuse component RT, 
 that have been modified with respect to Test 2 are the following: i) the discrete source has been
placed in the centre of the box and the box size has been increased by a factor 2, ii) the medium include hydrogen and helium,
iii) we evolve the simulation to $t=500$ Myr. Note that we do not expect in this case a full
equilibrium to be reached, given the complex interplay between hydrogen and helium recombination emission and their
ionization state. 

\begin{figure}
\begin{center}
\includegraphics[totalheight=0.27\textheight]{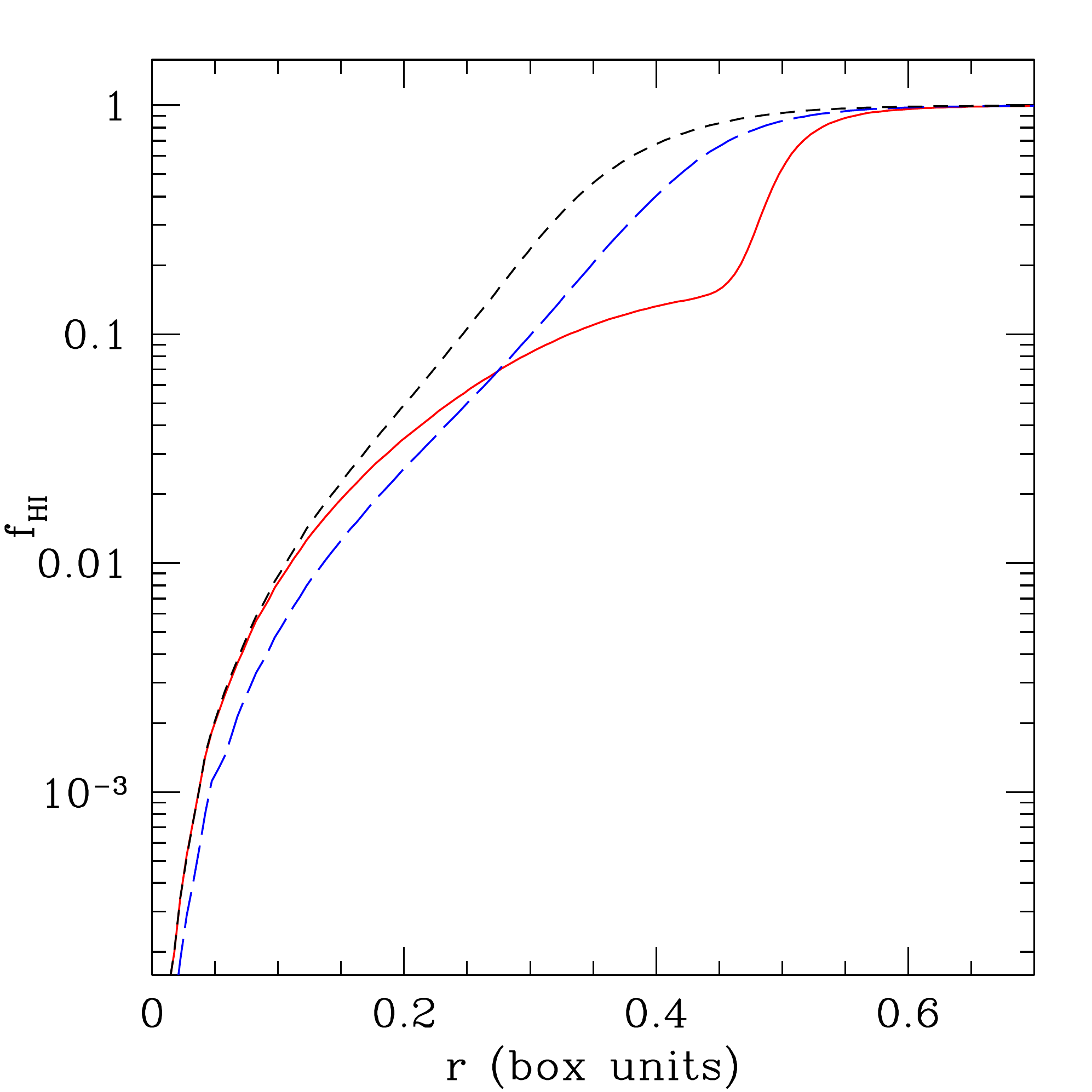}
\caption[Test 8: diffusion emission]
{Test 8: (diffuse emission from HII, HeII and HeIII) 
HI fraction profile at $t=500$ Myr for Case A (short dashed black line), Case B (long dashed blue line) 
approximations and including the full radiative transfer of ionizing diffuse emission 
from HII, HeII and HeIII recombinations (solid red line).     }
\label{HIfracDiff_T2b}
\end{center}
\end{figure}

\begin{figure}
\begin{center}
\includegraphics[totalheight=0.27\textheight]{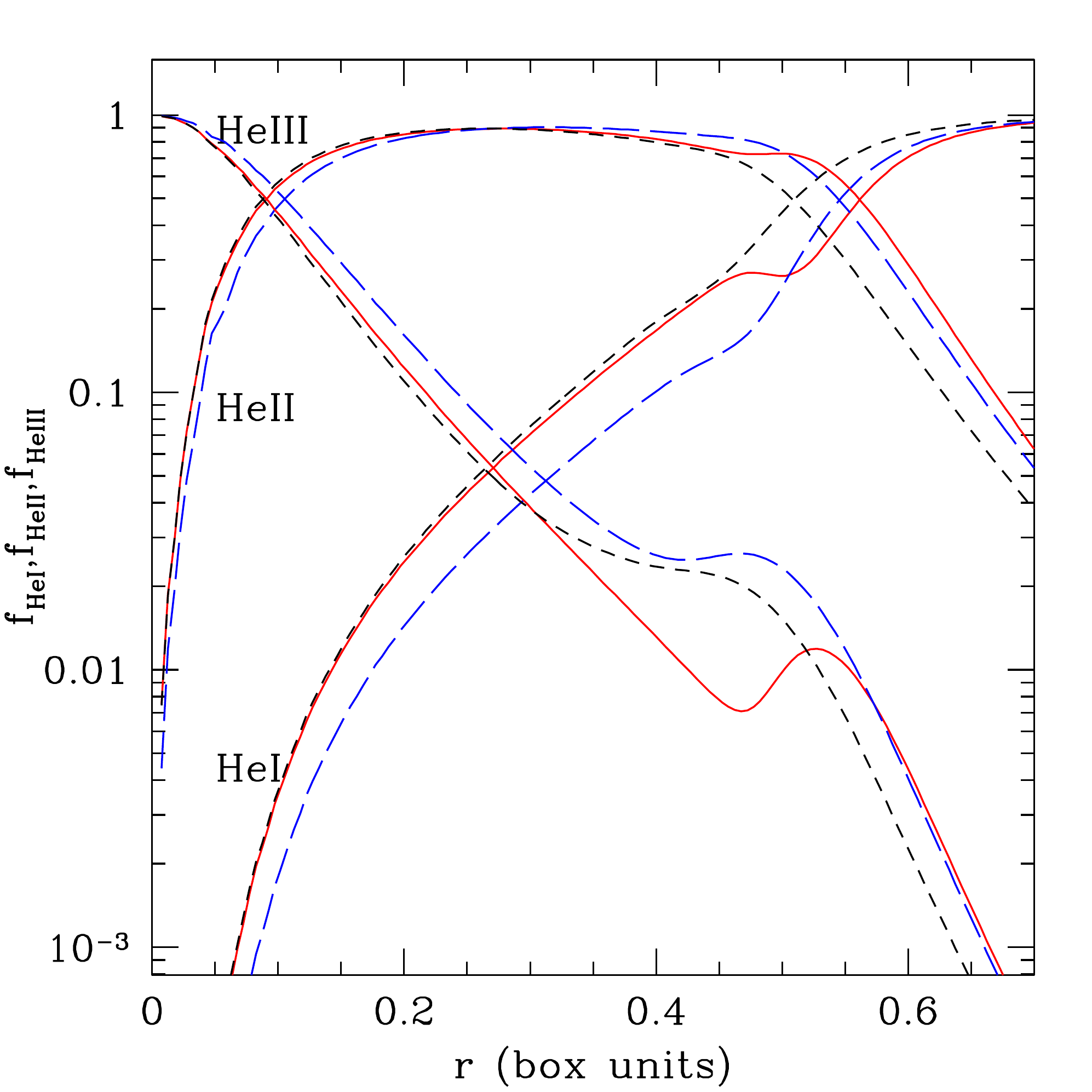}
\caption[Test 8: diffusion emission]
{Test 8: (diffuse emission from HII, HeII and HeIII) 
HeI, HeII and HeIII fraction profiles at $t=500$ Myr for Case A (short dashed black lines), ``on-the-spot'' Case B (long dashed blue lines) 
approximations and including the full radiative transfer of ionizing diffuse emission 
from HII, HeII and HeIII recombinations (solid red lines).
}
\label{HefracDiff_T2b}
\end{center}
\end{figure}

 In Figure \ref{HIfracDiff_T2b}, we present the HI profile obtained by the full RT of diffuse radiation (red solid line) in comparison
with the Case A approximation (black, short-dashed line) and the OTS Case B (blue, long-dashed line) at $t=500$ Myr. In Figure \ref{HefracDiff_T2b},
we show the analogous profiles for the HeI, HeII and HeIII fractions (see labels in the figure). As evident from Figure  \ref{HIfracDiff_T2b},
there is now an excess of HI photoionizations with respect to Case B, that leads to a larger Str\"omgren sphere size. This excess is due
to diffuse photons produced by helium recombinations, in particular the Balmer continuum (plus Ly$\alpha$ and two-photon continuum) 
produced in the HeIII region, that alters the total photon budget available for the photoionization of HI. In effect, a fraction of the 
high-energy, HeII-ionizing photons from the central source are ``converted'' via HeIII recombinations to lower energy, HI-ionizing photons 
(and also HeI-ionizing photons; indeed, a similar but less conspicuous effect is also visible for the HeI fractions in Figure \ref{HefracDiff_T2b}). 
Therefore, assuming Case B for hydrogen and helium actually results in the loss of this important component of the photon budget.
This enlightens the importance of including HeII and HeIII recombination emission also for studies that are only interested
in the properties of the hydrogen component.

\section{Discussion and conclusions}

We have presented a new, three-dimensional radiative transfer code, called \cnamens, 
based on an adaptive Monte Carlo ray-tracing scheme. 
 The algorithm has been specifically developed to efficiently resolve the small scales associated
with the Ionization-fronts within large, cosmological simulations, with the help of 
an Adaptive Mesh Refinement (AMR) scheme combined with a new ``cell-by-cell Monte Carlo'' approach:
rays are casted separately from selected cells within the computational box to the sources with a method
that ensures their correct solid angle distribution.

 This method has several advantages with respect to a classical Monte Carlo ray-tracing scheme 
\begin{footnote}{where a series of rays is uniformly casted around a source to obtain the correct solid angle
distribution.}\end{footnote}, in particular for multi-mesh domains. Indeed, in a classical Monte Carlo,
the number of rays to cast is determined by the smallest, most distant cells to the source, i.e.
from the (few) cells at the highest refinement level, oversampling the rest of the box.
In \cnamens, the number of rays is proportional to the number of cells to be evolved
during the current time-step, i.e. the fast evolving cells typically located within the I-front. 
This translates into a huge gain in computational speed and efficiency, 
given the small fraction of the computational volume occupied by the I-front. 
 Moreover, we are now able to choose a local criterion
that adaptively determines the number of rays needed in a particular cell, e.g. set by the
convergence of the radiation field. This ensures to obtain the required accuracy in the prediction
of the temperature or ionization state of the gas in a very efficient way: concentrating the
computational efforts only where actually needed.

 In the same spirit of hydrodynamical AMR codes, \cname is also able to increase the spatial resolution
where required, adaptively refining the mesh in correspondence of the Ionization-fronts. The implemented algorithm is
based on the original patch-based AMR method of Berger \& Rigoutsos (1991).

 \cname is able to trace the Ionization-fronts from multiple sources as well as the
diffuse ionizing radiation produced by hydrogen and helium recombinations with a
multi-frequency approach. The
time-evolution of six different species (HI, HII, HeI, HeII, HeIII, e) and 
the gas temperature is followed with a time-dependent, non-equilibrium chemistry solver 
based on an implicit Runge-Kutta scheme of variable, adaptive order.

 We performed all the four tests present in the radiative transfer code comparison project 
of Iliev et al. 2006, plus four additional, new tests aimed to substantiate and show the new
characteristic of \cname. The first four tests include the correct tracking of the I-front
in homogeneous (Test 1) and in-homogeneous density fields with multiple sources (Test 4), 
I-front trapping behind a dense clump (Test 3), and the effect of photo-heating 
on the gas temperature state (Test 2). These tests, like in the original I06 work,
have been performed on a single, static mesh, without emission from atomic recombinations
 in order to better compare our results with the other codes. 

 \cname results are in very good agreement with the majority of the
other codes present in I06. In particular, thanks to the new adaptive algorithm, we have shown 
(Test 1) that the recovered I-front thickness and structure does not suffer from any diffusive effects
and numerical broadening typical of other Monte Carlo codes (given their poor sampling 
at large distances from the source). The same is true when the gas temperature  
is examined (Test 2). 

 The second set of tests (Test 5 to Test 8) shows the ability of \cname to deal with both 
static (Test 5) and adaptively evolving (Test 6) multi-mesh structures, and with
diffuse radiation produced by hydrogen (Test 7) and helium (Test 8) recombinations. 
In Test 5, we reproduce the same results of the classical Str\"omgren sphere situation (Test 1) with
a nested hierarchy of meshes at different refinement levels, showing that no spurious
effects are introduced by the multi-mesh structure.

Most importantly, we have shown in Test 6 that \cname is able to fully resolve an expanding I-front
from a bright source (e.g., a quasar) within a large, cosmological box (100 comoving Mpc size),
thanks to an adaptively evolving mesh with several levels of refinement. The ability of resolving the
I-front on cosmological scales
is fundamental for a large range of applications. For instance, recovering the correct gas temperature within the I-front and the QSO bubble is 
important, e.g., for the study of the reionization epoch with the near-zone Ly$\alpha$ forest (see e.g., Bolton et al. 2010) 
or with the Ly$\alpha$ emission generated within the I-front (Cantalupo et al. 2008).
In Test 6, we have demonstrated that 
a proper treatment of the spectral hardening inside a resolved I-front may result in a substantial 
increase of the gas temperature ($\Delta T \sim10^4$ K) within the \emph{whole} HII region surrounding the bright, central
source. 

 The effect of diffuse radiation generated by hydrogen recombinations on the classic Str\"omgren sphere
case (with monochromatic radiation) has been presented in Test 7. Here, we have verified that our treatment of diffuse radiation
produces the same Str\"omgren sphere size with respect to the Case B approximation, in agreement
with photon conservation. Moreover, in accordance with the analytical prediction by Ritzerveld (2005), 
we have verified that the diffuse field becomes the dominant component at a distance from the source 
corresponding to 87\% of the Str\"omgren radius. The neutral fraction profile in the inner part of the
HII region follows the same profile obtained with Case A approximation, while in the outer part
the gas is more ionized and the I-front narrower than the result obtained with the widely used, Case B
approximation. In Test 8, we have also considered the effect of HeI and HeII recombination radiation
on the hydrogen and helium ionization state. We have found that HeII diffuse radiation, especially
the HI-ionizing, HeII Balmer continuum (together with HeII two-photon continuum and Ly$\alpha$ emission)
may have an important effect also on the hydrogen ionization state, substantially increasing the Str\"omgren sphere
size. 

 At present, \cname is not yet coupled with hydro-dynamics but is 
able to post-process the output of three different hydrodynamical codes, both grid-based, 
e.g., \texttt{RAMSES} (Teyssier 2002), \texttt{CHARM} (Miniati \& Colella 2007) and particle-based, e.g. \texttt{GADGET} (Springel 2005). 
These outputs are efficiently converted to \cname patch-based structures with a fast algorithm based
on the clustering method of Berger \& Rigoustous (1991), also used into the AMR module. 
The default format is very similar to the
widely used \texttt{CHOMBO HDF5} structure\begin{footnote}{https://seesar.lbl.gov/anag/chombo}\end{footnote}, 
allowing \cname outputs to be directly visualized with the most
recent, high-performance visualization packages (e.g., \texttt{VISIT}\begin{footnote}{https://wci.llnl.gov/codes/visit}\end{footnote}). 
Currently, \cname is efficiently parallelized
with OpenMP. All the results presented in this paper have been obtained with a few hours of computational time on a 8-core Intel-Xeon workstation.

\section*{Acknowledgments}

SC thanks Francesco Miniati for useful discussions and Martin Haehnelt for comments on an earlier version of this manuscript. 
CP acknowledges the role of Piero Madau who nearly 10 years ago sensitised him to the cosmological radiative-transfer problem and followed 
the development of some unpublished algorithms. Part of the figures presented 
in this paper have been realized with the visualization software VISIT, we are grateful to LLNL to have made public this powerful tool.

\appendix

\section{Analytical approximation for the solid angle of a cubic cell}

 Starting from the explicit form of the solid angle for each face of a cubic cell, we have found the following approximation for the total cell solid
angle as seen by any source inside (or outside) the computational box:

\begin{equation}\label{AnSA}
\begin{split}
\frac{\Omega_{cell}}{\mathrm{sr}} = & \displaystyle\sum_{n=1,2,3} \ \displaystyle\sum_{\delta=-1,1} \mathrm{MAX} \ \bigg[\ 0 \ , \ 
                \displaystyle\sum_{i=1}^8 \frac{(\Delta_{in}-\delta)}{2} \times \\ 
              & \frac{\lvert \epsilon_{pqn}\rvert (\epsilon_{pqn}+1)}{2} 
                \arctan \bigg(\frac{\Delta_{ip}\Delta_{iq}\Delta_{in}x_{ip}x_{iq}}{x_{in}\sqrt{x_{ip}^2+x_{iq}^2+x_{in}^2}}\bigg)  \bigg] \ ,
\end{split}
\end{equation}
where $\epsilon_{pqn}$ is the three-dimensional Levi-Civita symbol (note the implicit sum over the indices $p$ and $q$),
\begin{equation}
x_{ij}=J_{8,1}\cdot(x_{j}^{cell}-x_{j}^{source})+\frac{l}{2}\Delta_{ij} 
\end{equation}
is the $8\times3$ matrix containing the cell vertexes coordinates (translated to the Cartesian system with the source at the origin), $x_{j}^{cell}$ and
$x_{j}^{source}$ are, respectively, the cell centre and the source coordinates,
$l$ is the linear cell size, $J_{8,1}$ is the $8\times1$ unit matrix (i.e., the matrix of ones), and
$\Delta_{i,j}$ is the following $8\times3$ matrix:
 
\begin{center}
$
\Delta=
\begin{bmatrix}
-1 & -1 & -1 \\
\ \ 1  & -1 & -1 \\
-1 &\ \  1 & -1 \\
\ \ 1  &\ \  1 & -1 \\
-1 & -1 &\ \  1 \\
\ \ 1  & -1 &\ \  1 \\
-1 &\ \  1 &\ \ 1 \\
\ \ 1  &\ \  1 &\ \ 1 \\
\end{bmatrix}
$
.
\end{center}

All lengths and coordinates are in box units.
Note that the presence of the factors containing $\Delta_{ij}$ and $\epsilon_{pqn}$ reduce the total number of terms in the overall sum to four per cell face. 
The function $\mathrm{MAX}$ removes from the sum the solid angle of a face \emph{invisible} to the source, since this has a negative sign. 

\begin{figure}
\begin{center}
\includegraphics[totalheight=0.27\textheight]{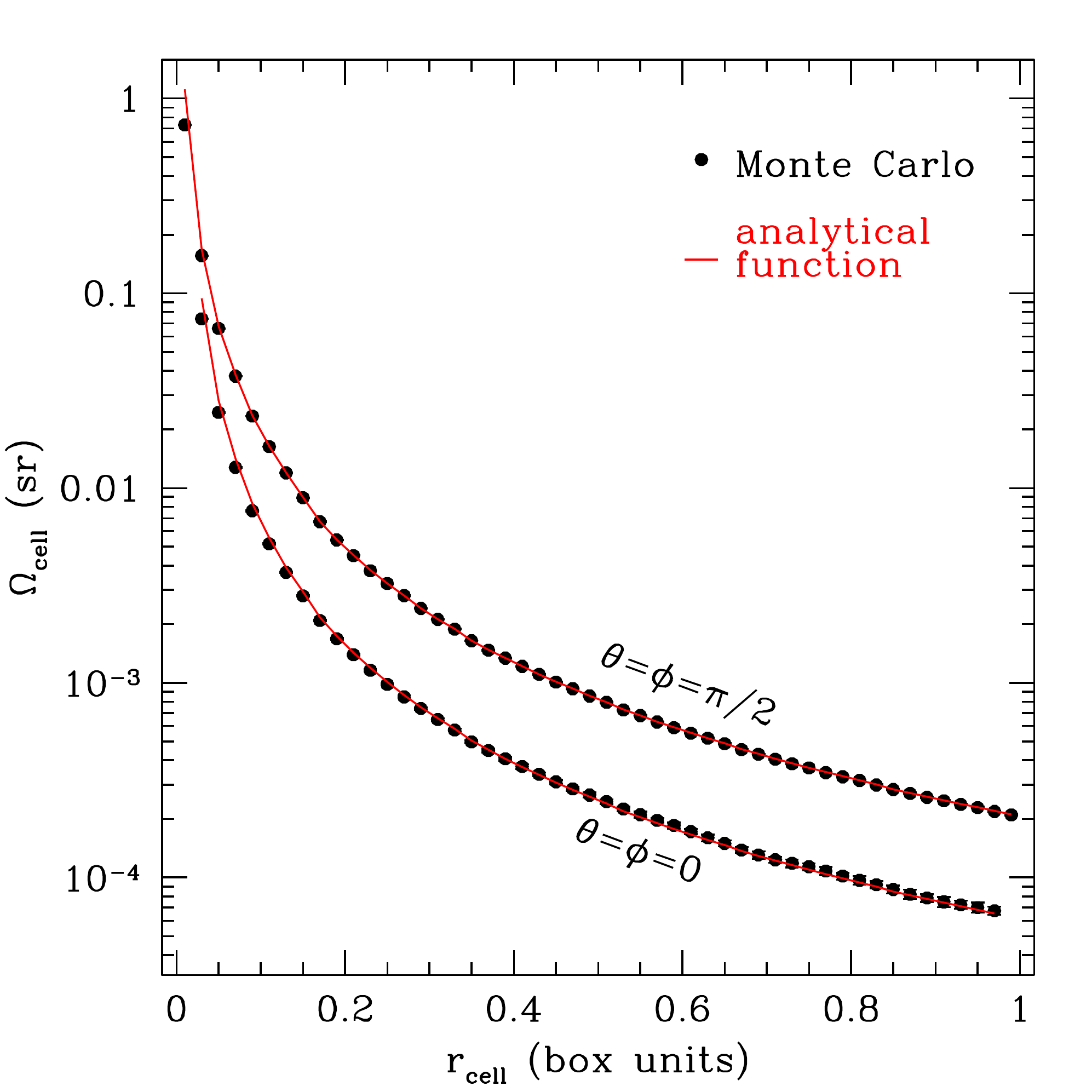}
\caption[]
{Solid angle test: comparison between the recovered cell solid angle by a full Monte Carlo simulation (black circles) and our analytical solution (red solid lines)
in a $128^3$ grid. The solid angle is shown as a function of the distance from the source, as seen by two different angular directions (see labels).}
\label{SAtest}
\end{center}
\end{figure}

 In Figure \ref{SAtest}, we show the comparison between the recovered cell solid angle
obtained by a full Monte Carlo simulation with $128^3$ cubic cells and $10^8$ rays (black circles),
and the above analytical approximation (red solid line). The solid angle
is shown as a function of the distance from the source, placed at $(l/2,l/2,l/2)$,
and for two different directions: parallel to the $x$-axis (i.e., normal to two cell faces; lower line) 
and along the line of sight that connects the source to the box vertex $(1,1,1)$ (upper line).
Note that these two directions correspond to the minimum and maximum possible solid angle of a cubic cell at a 
given distance from the source. 
The analytical approximation is in very good agreement with the Monte Carlo results.


\section{Test 6 for a source with soft spectrum}

\begin{figure}
\begin{center}
\includegraphics[totalheight=0.25\textheight]{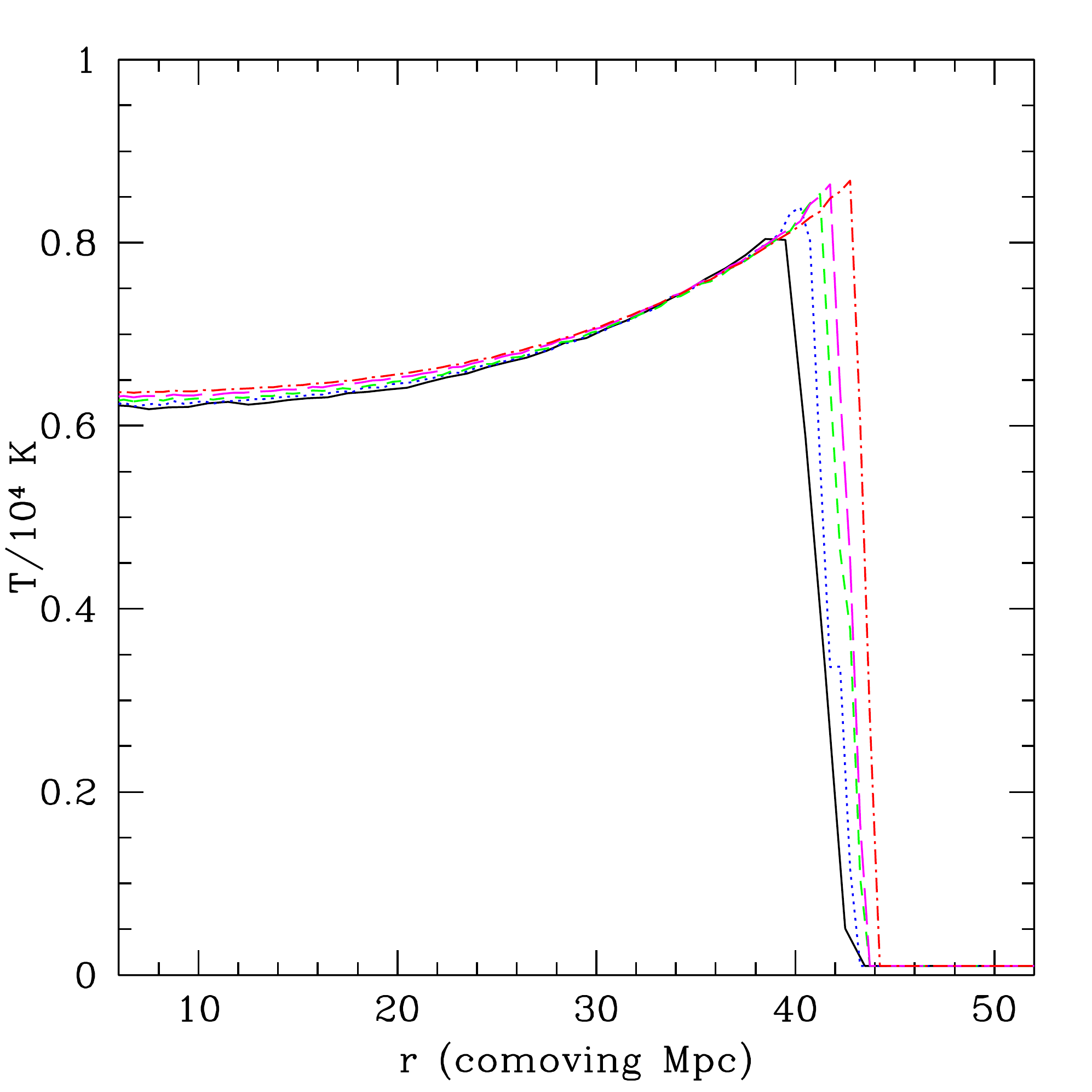}
\caption[]
{Same as Figure 18 (Test 6) but for a source with a soft spectrum (a black-body with $T=2\times 10^4$ K). In this case, spectral hardening
is minimal and different maximum levels of refinement produce now very similar temperatures inside the HII region. This demonstrates
that numerical effects are not substantially affecting the results of Test 6.}
\label{T6B}
\end{center}
\end{figure}

In Test 6, we have shown that the spectral hardening inside a resolved I-front may result in a substantial 
increase of the gas temperature ($\Delta T \sim10^4$ K) within the \emph{whole} HII region surrounding a bright quasar.
To better demonstrate that this result is not affected by 
numerical effects due to the different grid resolutions, we have repeated Test 6 assuming a source
with similar ionizing rate but with a soft spectrum (a black-body with $T=2\times 10^4$ K).
In this case, spectral hardening is minimal and therefore
we do not expect that increasing the resolution on the I-front should substantially change the gas temperature as in Test 6.  
As we show in Figure \ref{T6B} (line colors and style have the same meaning as in Figure 18), this is indeed the case: 
different maximum levels of refinement produce now very similar temperatures inside the HII region showing
that numerical effects are not important. 
Note that the (larger) shift in the I-front position in Figure \ref{T6B} (with respect to Figure 18) is due to the fact that,
at the lowest resolution, cells are very optically thick to radiation with frequencies right above the ionization threshold
(the vast majority, given the soft spectrum assumed here). This slightly alters photon-conservation, which holds to high accuracy 
when the front is resolved into less optically thick elements, as shown in Test 1.

\label{lastpage}

\end{document}